\documentclass[prb,twocolumn,showpacs,preprintnumbers,amsmath,amssymb]{revtex4-1}
\usepackage{graphicx}% Include figure files
\usepackage{epstopdf}
\usepackage{color}
\usepackage{bbold}

\begin{document}
%\title{Low temperature phase diagram of voltage-gated oxide interfaces with strong Rashba coupling} 
\title{Phase diagrams of voltage-gated oxide interfaces with strong Rashba coupling} 
\author{D. Bucheli$^1$, M. Grilli$^{1,2}$, F. Peronaci$^3$, G. Seibold$^4$, and S. Caprara$^{1,2}$} 
\affiliation{
$^1$Dipartimento di Fisica Universit\`a di Roma Sapienza, piazzale Aldo Moro 5, 
I-00185 Roma, Italy\\
$^2$ISC-CNR and Consorzio Nazionale Interuniversitario per le Scienze Fisiche della 
Materia, Unit\`a di Roma Sapienza, Italy\\
$^3$Scuola Internazionale Superiore di Studi Avanzati (SISSA),
via Bonomea 265, I-34136 Trieste, Italy\\
$^4$Institut f\"ur Physik, BTU Cottbus-Senftenberg, PBox 101344, 03013 Cottbus, 
Germany}
%\author{} 
%\affiliation{}   
\begin{abstract} 
We propose a model for the two-dimensional electron gas formed at the interface of oxide 
heterostructures that includes a Rashba spin-orbit coupling proportional to an electric field 
oriented perpendicularly to the interface. Taking into account the  electron density dependence 
of this electric field confining the electron gas at the interface, we report the occurrence of 
a phase separation instability (signaled by a negative compressibility) for realistic values of 
the spin-orbit coupling and of the electronic band-structure parameters at zero temperature. We 
extend the analysis to finite temperatures and in the presence of an in-plane magnetic field, 
thereby obtaining two phase diagrams which exhibit a phase separation dome. By varying the gating 
potential the phase separation dome may shrink and vanish at zero temperature into a quantum 
critical point where the charge fluctuates dynamically. Similarly the phase separation may be 
spoiled by a planar magnetic field even at zero temperature leading to a line of quantum critical 
points.
\end{abstract}   
\date{\today} 
\pacs{71.70.Ej,73.20.-r,73.43.Nq,74.81.-g} 
\maketitle 
   
\section{Introduction}   
The observation of a two-dimensional (2D) metallic state at the interface of two insulating oxides 
LaAlO$_3$/SrTiO$_3$ (LAO/STO) \cite{ohtomo,Mannhart:2008uj,Mannhart:2010ha,Hwang:2012nm} has brought 
forward a novel class of high-mobility electron gases (EGs) which proves to be important in both 
theoretical and practical prospects. The occurrence of superconductivity in this 2DEG\cite{reyren,triscone} 
and in similar heterostructures such as LaTiO$_3$/SrTiO$_3$ (LTO/STO),\cite{espci1,espci2} with 
the possibility to control the transition by voltage gating, has further attracted great attention. 
There is however increasing evidence that inhomogeneity plays a relevant role in these systems. Not only 
is the large width of the superconducting transition measured in transport experiments a clear indication 
of charge inhomogeneity,\cite{CGBC,BCCG,caprara} 
but also magnetometry,\cite{ariando,luli,bert,metha1,metha2,bert2012} tunneling,\cite{ristic} and 
piezoforce spectroscopy\cite{feng_bi} report inhomogeneities on a submicrometric scale. 
Extrinsic mechanisms, like impurities and defects, surely introduce inhomogeneities.\cite{bristowe} 
However, the recent discovery of negative compressibility in a low filling regime\cite{mannhart} 
provides, in our opinion, a clear indication that \emph{intrinsic} mechanisms (i.e., effective 
electron-electron attractions) are present, which may render these 2DEGs inhomogeneous by phase 
separation even in a perfectly clean and expectedly homogeneous system. Moreover, even if such 
mechanisms were not strong enough to drive the EG unstable, they would still increase the 
charge susceptibility, reinforcing the extrinsic mechanisms. 

Several mechanisms providing negative contributions to the electronic compressibility are 
known,\cite{eisenstein} like the exchange term of the repulsive electron-electron interaction or 
the confinement mechanism forcing the EG at the interface. The former mechanism is only active at 
very low densities (even for semiconducting systems) up to a threshold density $n_c$, which is 
further reduced by large values of the dielectric constant. Therefore, this mechanism is surely 
not effective at the densities relevant for the metallic regime of the 2DEG in LXO/STO interfaces 
(henceforth, the symbol LXO stands for LAO or LTO, whenever we generically refer to both), where the 
typical electron density is quite large, of order of $10^{13}$ electrons per cell, and the STO 
dielectric constant may reach very high values, of order of $10^4$. On the other hand, concerning 
the latter mechanism, the self-consistent solution of the Schr\"odinger and Poisson equations
relating the electronic wavefunctions and the electric potential arising from external potential and
electronic density itself usually shows\cite{eisenstein} that the EG becomes more compressible once its 
finite transverse confinement is taken into account. Also in this case the large dielectric constant 
of the material hosting the EG favors its ``softening'', leading to an increased compressibility. 
Whether and under which conditions this mechanism may lead alone to a negative compressibility and 
to phase separation is presently under investigation.\cite{scopigno} In this general framework, looking 
for possible realistic sources of intrinsic mechanisms of phase separation, we recently proposed a 
possible explanation for the occurrence of an inhomogeneous  electronic phase, which is based on a 
coupling between the electron orbital degree of freedom and its spin.\cite{CPG} The confinement of the 
2DEG at the interface restricts the electronic motion to the plane parallel to the interface, while 
the electric field confining the electrons is oriented perpendicularly. In this configuration, the 
interaction between the moving electron and the electric field gives rise to the so-called Rashba 
spin-orbit coupling (RSOC);\cite{rsoc} in simple terms, the moving electron perceives the 
electric field as a magnetic field in its rest frame, which then couples to its spin. As it will 
be explained in detail below, the key point of our argument is the dual role assumed by the confining 
electric field; the field is proportional to the electron density and controls at the same time the 
RSOC. This leads to a density-dependent RSOC (via the electric field). Since the band structure depends 
on the strength of the coupling, the system acquires a \emph{non-rigid band structure} which evolves as 
a function of the electron density. This non-rigidity gives rise to a 2DEG which may differ strongly 
from a standard metallic system with rigid bands. For instance, the chemical potential $\mu$ in the 
latter case is an increasing function of the electron density $n$ and the compressibility 
$\kappa\equiv\partial n/\partial\mu$ is always positive. This statement holds no longer true for a 
system with non-rigid bands; the 2DEG may be in a (thermodynamically unstable) phase with a 
negative compressibility and separate into two sub-phases with densities $n_1$ and $n_2$.

The analysis of the precise conditions leading to such an unstable phase will be the focus of this paper, 
which is organized as follows. In section \ref{Sect_Model} we present in detail our model used for the 
description of the 2DEG at the interface. We study the model at zero and low temperatures as well as zero 
and small in-plane magnetic field and report the obtained results in section 
\ref{Sect_Results}-\ref{magn-field}. Our concluding remarks are given in section \ref{Sect_Conclusion}.

\section{Model}\label{Sect_Model}   
\subsection{Origin of the 2DEG}
\begin{figure}[h]
\begin{center}
\includegraphics[scale=0.235]{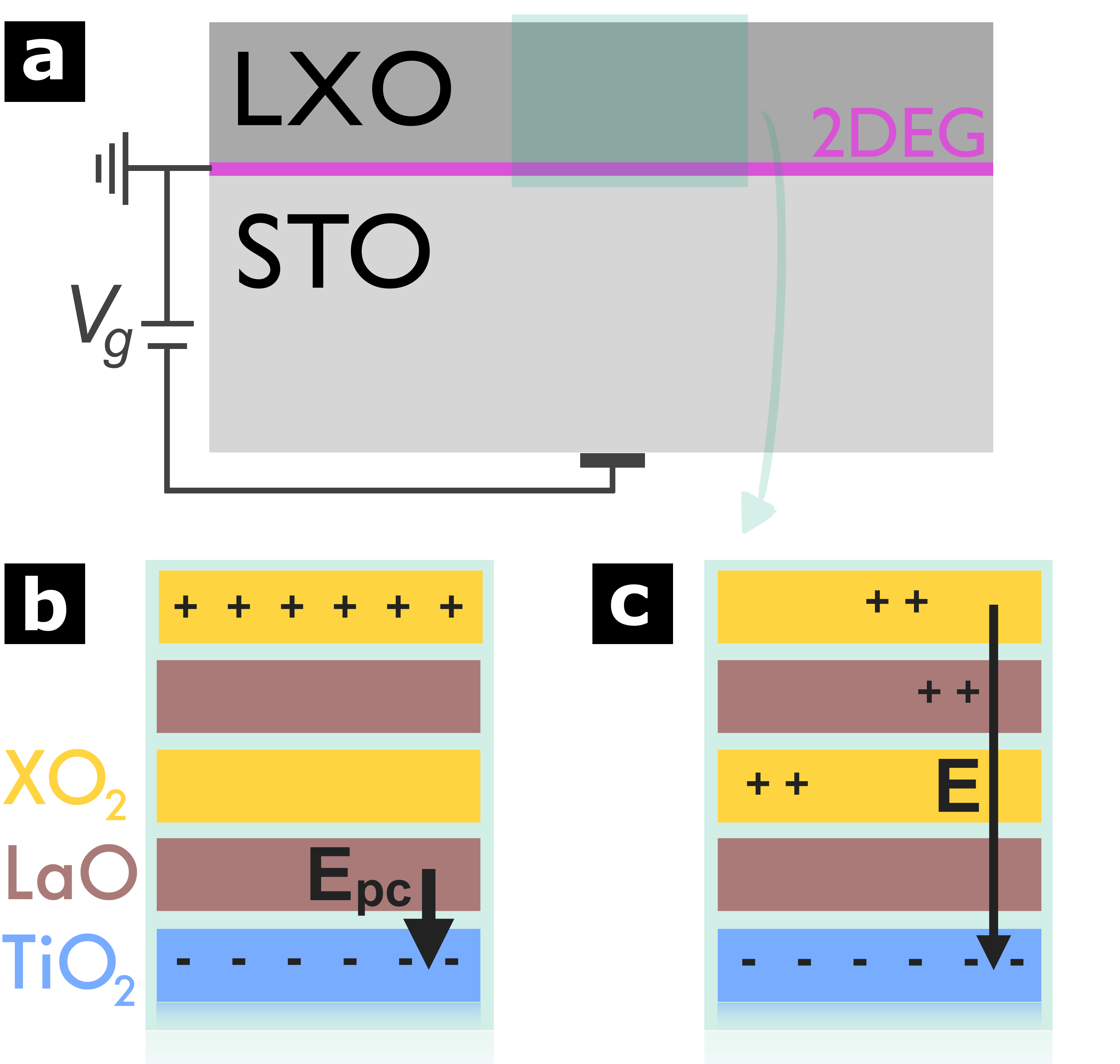}
\caption{(color online) 
(a) Schematic view of the LXO/STO interface in the presence of a gate potential $V_g$. The 2DEG created 
at the interface resides in the STO conduction bands. 
(b) Illustration of electronic reconstruction due to polarity catastrophe in an n-type interface at zero 
gating. $E_{pc}$ is the associated electric field. (c) Illustration of the metallic interface due to 
oxygen vacancies.}
\label{PolarCat}
\end{center}
\end{figure}

%%%   P O L A R     C A T A S T R O P H E   %%%
The prototype systems we have in mind are oxide heterostructures formed by an STO bulk with a thin LXO 
film (of the order of a few unit cells) on top [Fig. \ref{PolarCat}(a)]. The LXO film consists of 
alternating \emph{polar} layers of (LaO)$^+$ and (XO$_2$)$^-$, with X$=$Al,Ti, whereas the STO bulk 
is formed by \emph{non-polar} layers of TiO$_2$ and SrO. This difference in polarity leads to a 
charge discontinuity at the interface. Depending on the termination, two different types of interface 
may arise: XO$_2$/LaO/TiO$_2$, called \emph{n-type}, and XO$_2$/SrO/TiO$_2$, 
called \emph{p-type}. In both cases, the charge discontinuity at the interface leads to a build-up of 
the electrical potential in the LXO film which increases monotonically with the number of layers and 
would diverge for an infinite number of layers. Due to this (theoretical) divergence, the scenario is 
termed \emph{polarity catastrophe}.\cite{polarity,popovic,bands1,bands2,bands3} According to theory, 
the system avoids this unphysical limit through a redistribution of the charges at the interface.
By adding half an electron per unit cell to the interfacial TiO$_2$ layer coming from the uppermost XO$_2$ 
layer (n-type), or by removing half an electron from the interfacial SrO plane in the form of oxygen 
vacancies (p-type), a small dipole arises that causes the electric field to oscillate about zero and the 
potential to remain bounded. While n-type interfaces exhibit a metallic behavior, experiments show that 
p-type interfaces are insulating. The experiments further report a smaller transferred amount of charge 
of order $n_0\sim 0.02$ electrons per unit cell (el./u.c.),\cite{hirayama,parkmillis} instead of 
the theoretical $0.5$\,el./u.c. (see Fig. \ref{PolarCat}(b) for an illustration). These charges create 
a strong electric field, 
\[
E_{pc}=\nu\,n_0,\quad\text{with}\:\:\:\nu=\frac{e}{\epsilon_0\epsilon_ra^2},
\]
where $\epsilon_r\sim 20$ is the dielectric constant of LXO and $a=3.905\times10^{-10}$\,m is the 
linear size of the STO unit cell.

The polarity catastrophe is not the only mechanism leading to the formation of a 2DEG at the 
LXO/STO interface. The gas can be also formed as a consequence of oxygen vacancies (mostly located in 
LXO, which is interposed between LTO and vacuum), that act as electron dopants, each oxygen vacancy 
nominally providing two electrons [see Fig. \ref{PolarCat}(c)]. These additional
electrons then are confined at the interface by the self-consistent potential due to the other 
electrostatic fields (e.g., bulk fields or gates) and the electronic density itself. Solving the 
Schr\"odinger and Poisson equations one finds that this confining potential generates an electric 
field perpendicular to the interface which is roughly proportional to the density of the confined 
electrons. Therefore also oxygen vacancies eventually produce an interface field, which is strictly 
related to the local density of the 2DEG, similarly to the outcome of the polarity catastrophe. The 
orders of magnitude, as expected in the presence of similar confined densities, should be comparable. 

In our model, we focus on n-type interfaces. For the sake of concreteness, but without being 
necessarily tied to this mechanism, we consider here electronic charges due to the polarity 
catastrophe and explore a range of values of $n_0$ up to $0.08$\,el/u.c. 
(corresponding to an electric field $E_{pc}\simeq 5\times10^8$\,V\,m$^{-1}$), in which,
as we will show, the system may undergo electronic phase separation. 

%%%   E F F E C T    O F    G A T I N G   %%%
\subsection{Effect of Gate Voltage}
\begin{figure}[h]
\begin{center}
\includegraphics[scale=0.23]{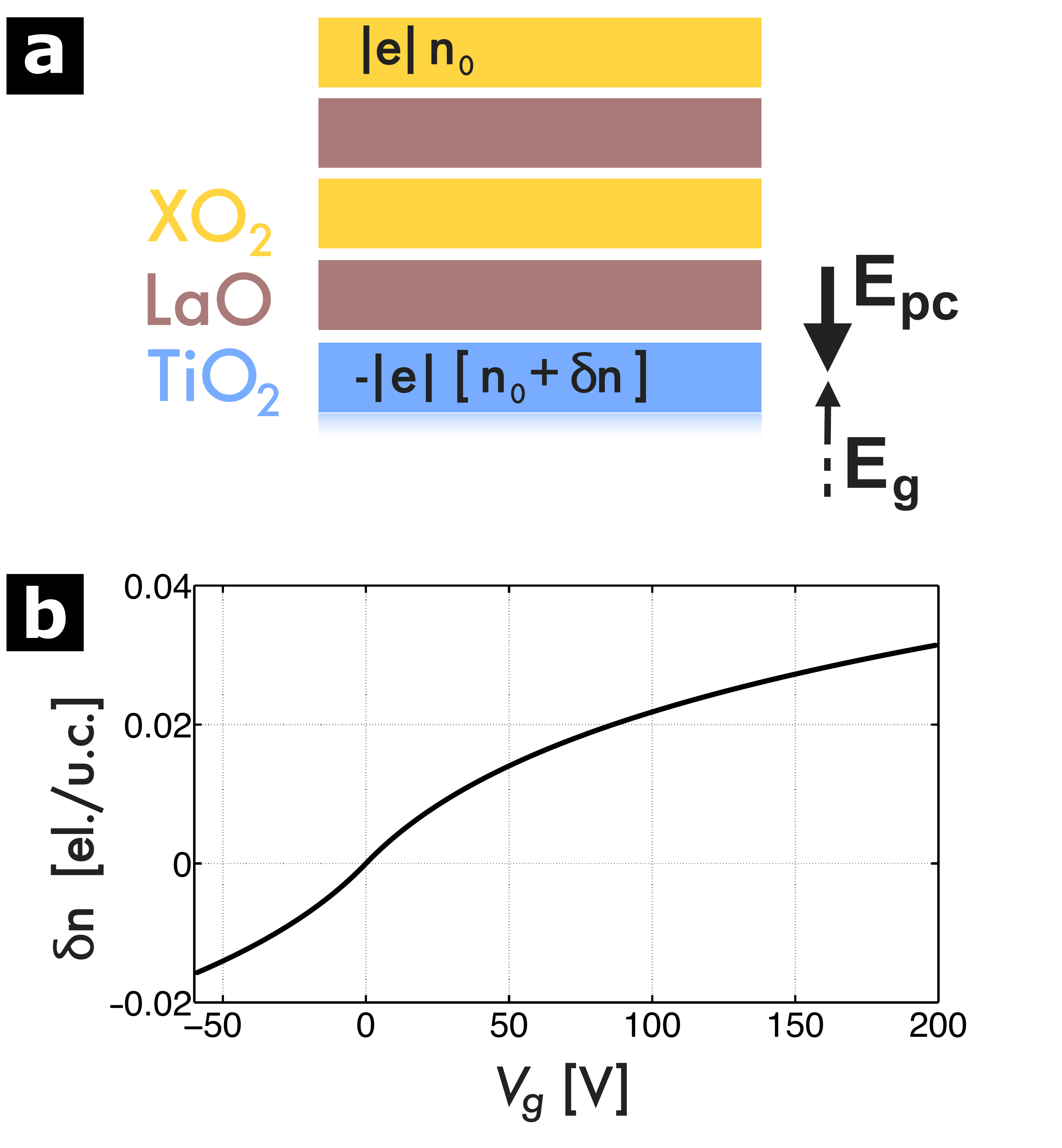}
\caption{(color online) 
(a)  Illustration of electronic reconstruction due to polarity catastrophe in an n-type interface at 
finite gating.
$n_0$ is the density of the electrons transferred to the interface by the polarity catastrophe giving 
rise to the $E_{pc}$ electric field. 
$\delta n$ is the electron density tuned by the gating field $E_g$, with $E_g\ll E_{pc}$. (b) 
Field-effect charge modulation in the STO bulk as a function of gate voltage. A carrier density 
of $0.016$\,el./u.c. corresponds to $10^{13}$\,cm$^{-2}$.
}
\label{ExpSetup}
\end{center}
\end{figure}

As mentioned, the density of the 2DEG can be additionally increased (decreased) by the application of 
a positive (negative) gate voltage on the sample. Typically, the voltage is applied by back-gating to 
the STO bulk alone, while the interface is kept at 
$V_g=0$ V. Applying a voltage $V_g\sim100$\,V to a layer of thickness $d\sim5\times10^{-4}$\,m
yields an electric field $E_g=V_g/d\sim 2 \times10^5$\,V\,m$^{-1}$, which is orders of magnitude 
smaller than the electric field due to the polarity catastrophe, $E_{pc}$ [see Fig. \ref{ExpSetup} (a)].

In the low temperature region we are interested in, the dielectric constant of STO is a highly non-linear 
function of the electric field of the form\cite{espci2,Neville,copie}
\begin{equation}
\epsilon_{STO}(E_g)=\epsilon_{\infty}+\frac{1}{A+B|E_g|},
\label{epsSTO}
\end{equation}
where $\epsilon_{\infty}=300$ is the saturation value for very high field, $A\sim4\times10^{-5}$ and 
$B\sim5\times10^{-10}$\,mV$^{-1}$ are temperature-dependent parameters and $|E_g|$ is the amplitude 
of the electric field. Considering systems at zero or low temperature, where the
values of $A$ and $B$ change slowly, we take $\epsilon_{STO}$ to be temperature independent.
The integration of the capacitance per unit surface $\epsilon_0\epsilon_{STO}/d$ 
from zero voltage to $V_g$ yields the charge density $|e|\delta n(V_g)$, where $\delta n(V_g)$ is the 
particle density plotted in Fig. \ref{ExpSetup} (b).
In particular, the 2DEG can be greatly depleted tuning the density to $\delta n\sim-n_0$, while the total 
electric field $E=E_g+E_{pc}\approx E_{pc}$ stays large. We will see that this implies the persistence of 
a strong RSOC also in the negative voltage region.

%%%   B A N D    S T R U C T U R E   %%%
\subsection{SrTiO$_3$ Band Structure}
The mobile electrons at the interface reside in the STO t$_{2g}$ conduction bands originating from 
the $3d$ orbitals of Titanium. Experiments and theoretical calculations suggest that for the strictly 
2D case the t$_{2g}$ bands are split into an isotropic band $d_{xy}$ (with light mass $m_l$) separated 
by an energy $\Delta$ from two anisotropic bands $d_{xz}$ and $d_{yz}$ (each with a light mass $m_l$ 
and a heavy mass $m_h$).\cite{salluzzo,santander,meevesana,held,nayak} In addition, a finite extension 
of the gas perpendicular to the plane may give rise to multiple $d_{xy}$ (and $d_{xz},d_{yz}$) 
sub-bands. While the experiments performed at the STO/vacuum interface report masses of the order of 
$m_l\sim0.6m_0$ and $m_h\sim10-20\,m_0$ ($m_0$ the bare electron mass), recent density functional theory 
(DFT) calculations report smaller masses of the order $m_l\sim0.5m_0$ and $m_h\sim1.1m_0$. Experiments 
indicate a gap $\Delta$ of the order of $50$\,meV.\cite{salluzzo} DFT calculations for LAO/STO report 
$\Delta\sim250$\,meV, with subbands which lay closer (at about $50$\,meV) to the bottom of the 
heavier bands. In any case these values depend strongly on the details of the interface and smaller 
values are conceivable. Caviglia et al. obtain RSOC values of the order of $10^{-12}-10^{-11}$\,eV\,m 
by fitting magnetoresistance measurements with Maekawa-Fukuyama and Dyakonov-Perel theory with a single 
band of mass $3m_0$. Zhong et al. find couplings of the same order in the limit $\Delta\rightarrow0$.

In addition to the rather wide range of reported parameters, the precise form of the RSOC in the 
different bands remains an open issue. While in Ref. \onlinecite{triscone} the RSOC at the LAO/STO 
interface was successfully fitted assuming a linear in $k$ dependence for all gate voltages, the authors 
of Ref. \onlinecite{Kimura} found evidence of \emph{cubic} Rashba spin-splitting in the 2DEG formed at 
the STO/vacuum interface. In Ref. \onlinecite{nayak} the authors arrived at yet another conclusion, 
reporting that the induced spin-orbit coupling in the $d_{xz}$ and $d_{yz}$ bands is cubic in $k$, 
whereas the spin-orbit interaction in the lower band ($d_{xy}$), has a linear momentum dependence. 
Ref. \onlinecite{held} points out that the inconsistencies may be partially resolved by the fact that 
the two interfaces are different. More to the point, the authors show that a negative $\Delta$, as it 
may occur at the STO/vacuum interface, leads to a cubic-in-$k$ RSOC. However, the similarity of the 2DEG 
at the STO surface to those reported in STO-based heterostructures suggests that different forms 
of electron confinement at the surface of STO should lead to essentially the same 2DEG\cite{santander}.

In the present paper, we shall mainly consider a model with linear RSOC and an expedient 
band structure composed of one anisotropic band with mass $m_l=0.7m_0$ separated by $\Delta=50$\,meV 
from two anisotropic bands with masses $m_l=0.7m_0$ and $m_h=21m_0$, allowing for a simpler
analytical treatment. For completeness, in Appendix A we also 
report results obtained for different values of the parameters $m_l, m_h, \Delta$ and RSOC. In addition, 
in Appendix B, we show that the main conclusions of our work remain valid 
for more complex band structures arising from orbital mixing, for which an analytical determination 
of the bands in not possible. Indeed, the DFT calculations mentioned above elucidate how the 
Rashba splitting arises from the combined effects of atomic spin-orbit coupling and the interfacial 
electric field, so that the corresponding tight-binding Hamiltonian is non-diagonal in the orbitals. 
However, to perform analytic calculations of various quantities, that are impossible for the complex 
tight-binding band structure derived from DFT calculations, we adopt in this work  a 
simplified Hamiltonian of the form:
\begin{equation}
{\cal H}_{tot}=
 \left( \begin{array}{ccc}
{\cal H}_1 & 0 & 0 \\
0 & {\cal H}_2 + \Delta & 0 \\
0 & 0 & {\cal H}_3 + \Delta
 \end{array} \right)
 \label{acca}
\end{equation} 
with the single-orbital Hamiltonian ${\cal H}_i$:
\begin{equation}\label{H_Rashba}
{\cal H}_i=\left( \frac{\hbar^2 {k}_x^2}{2m_{x,i}}+\frac{ \hbar^2 {k}_y^2}{2m_{y,i}}\right)\sigma_0+\alpha\bigl[  {k}_y\sigma_x-  
{k}_x\sigma_y\bigr],
\end{equation}
where $\sigma_x$ and $\sigma_y$ are the standard Pauli matrices ($\sigma_0$ is the unity matrix) 
and $\alpha$ is the strength of the 
RSOC. The first band is isotropic, $m_{x,1}=m_{y,1}=m_l$, whereas the second and third bands
are anisotropic, $m_{x,2}=m_{y,3}=m_l$, $m_{y,2}=m_{x,3}=m_h$. The case of a $k$-cubic RSOC
is briefly discussed in Appendix C, where we show that also this modification of our model leads 
to very similar physical results.

Diagonalizing the Hamiltonian of Eq. (\ref{H_Rashba}) we obtain the following dispersion relations:
\begin{flalign}
&\text{isotropic band:}\nonumber \\ 
&\varepsilon^{\text{i}}_{\pm}(k)=\frac{\hbar^2k^2}{2m_l}\pm\alpha k,\;\text{with }k=\sqrt{k_x^2+k_y^2},
\label{disp_rel_iso}\\
&\text{ }\nonumber \\
&\text{anisotropic bands:}\nonumber \\
&\varepsilon^{\text{a}}_{\pm}(k_x,k_y)=\frac{\hbar^2k_x^2}{2m_x}+\frac{\hbar^2k_y^2}{2m_y}+\Delta
\pm\alpha k,
\label{disp_rel_aniso}
\end{flalign}
where $m_x=m_l(m_h)$, $m_y=m_h(m_l)$ for the second (third) band of Eq. (\ref{acca}).
Although an anisotropic RSOC could be considered for the anisotropic bands,\cite{CPG} for simplicity, we
assume here the same RSOC for all bands. Henceforth, the superscripts $^\text{i}$ and $^\text{a}$ label 
quantities defined for the isotropic and the anisotropic bands, respectively. 

Due to the RSOC the quantum number associated to the spin degree of freedom is no longer the spin itself 
but the so-called chirality, which for a given band and wave vector $k$ characterizes the orientation
of the eigenspinor which we label by $+$ and $-$. The resulting Fermi surface is formed by two circles 
(ellipses for the anisotropic bands) and shown in Fig. \ref{BandStructure_Bfield}. The peculiar spin 
structure will be important when we consider the effect of a magnetic field parallel to the interface. 

\begin{figure}[h]
\begin{center}
\includegraphics[width=1.0\linewidth]{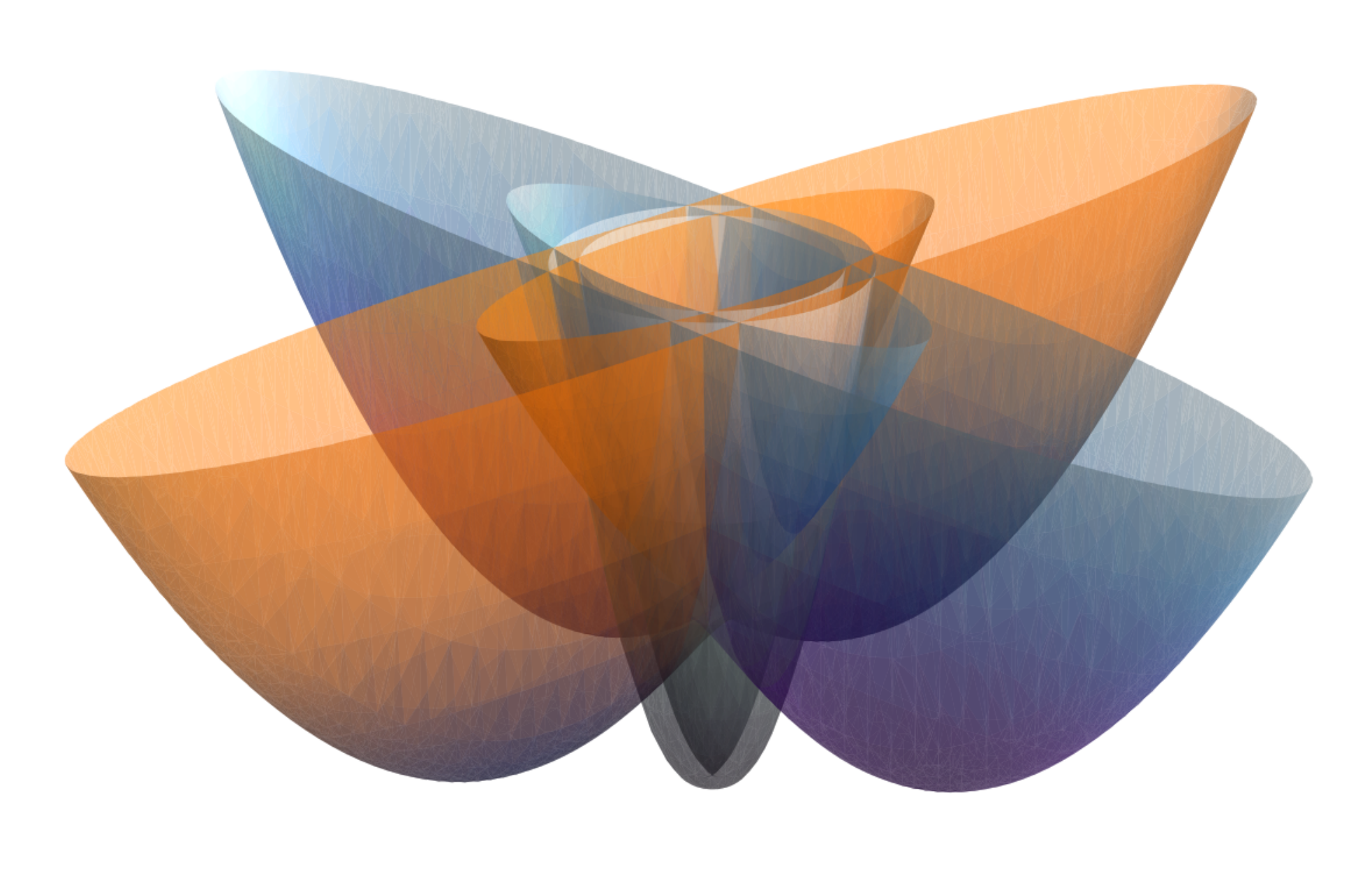}
\caption{(color online) Schematic view of the STO band structure formed by an isotropic Rashba band 
(grey) and two anisotropic bands (orange and blue). The isotropic band has a ring of minima at 
$k_0^{\text{i}}=m^*\alpha/\hbar^2$ while the anisotropic bands have each two minima at 
$k_0^{\text{a}}=\pm m_h\alpha/\hbar^2$, where $k_0^{\text{a}}$ 
is along the direction with the heavy mass $m_h$.}
\label{bandstructure}
\end{center}
\end{figure}

The anisotropic bands have each two minima at $k_0^{\text{a}}=\pm m_h\alpha/\hbar^2$ on the axis 
corresponding to the heavy mass. The separated minima become a ring of radius 
$k_0^{\text{i}}=m_l\alpha/\hbar^2$ in the isotropic band. A schematic view of the resulting band 
structure is given in Fig. \ref{bandstructure}.

We will see in the following how, due to the density-dependent Rashba coupling, the band structure, and 
in particular its (local) minima $\epsilon^{\text{i,a}}_0=\alpha k^{\text{i,a}}_0/2$ are functions of 
the electron density.

%%%   A L P H A (  n 0  )   %%%
\subsection{Field-dependent Rashba Coupling}
Concerning the dependence of the Rashba coupling on the electric field, in the absence of 
compelling first-principle calculations, we borrow its functional form 
from semiconductor physics, while the appearing parameters are inferred from LXO/STO experiments.
More precisely, we take\cite{libroSO,schapers}
\begin{equation}\label{kp}
\alpha\propto\bigg\langle\psi(z)\bigg|\frac{d}{dz}\biggl[\frac{1}{\bar{\varepsilon}(z)}-
\frac{1}{\bar{\varepsilon}(z)+\Delta^{SO}}\biggr]\bigg|\psi(z)\bigg\rangle,
\end{equation}
where $\psi(z)$ is the  electron wave function (for simplicity we here display only its $z$ dependence)
and $\bar{\varepsilon}(z)\equiv \varepsilon+V(z)+E_{gap}$, $\varepsilon$ is the subband energy relative
to the bulk conduction band, $V(z)$ is the band-bending potential, $E_{gap}$ is the band gap, and 
$\Delta^{SO}$ is a measure of the spin-orbit splitting within a Kane $\bf{k}\cdot\bf{p}$ 
approach.\cite{kane} Expanding for small values of $\Delta^{SO}$, taking the electric field 
$E\simeq E_{pc}$ uniform along $z$,\cite{notapotential} and
replacing $V(z)$ in the denominator by $E\,\overline{z}$ after the differentiation,
($\overline{z}$ being the width of the confining potential well),
Eq. (\ref{kp}) can be written in the simple form
\begin{equation}\label{alpha_E}
\alpha(E)\simeq\alpha(E_{pc})=\frac{\gamma E_{pc}}{\bigl(1+\beta E_{pc}\bigr)^3}, 
\end{equation}
which in terms of the polarity catastrophe electron density $n_0$ becomes
\begin{equation}\label{alpha_n0}
\alpha(n_0;V_g)\simeq\alpha(n_0)=\frac{\gamma\nu\:n_0}{\bigl(1+\beta\nu\:n_0\bigr)^3}.
\end{equation}
The expression (\ref{alpha_E}) has the standard linear behavior at small electric field, but saturates at 
$E\sim\bar{E}=(2\beta)^{-1}$ and then decreases at larger fields. This latter behavior is important 
to stabilize the system against an unphysical unbounded growth of the RSOC.
We define the parameters $\beta$ and $\gamma$ on purely phenomenological grounds. Experimental 
observations\cite{cavigliaPRL,fete,espciprivcom} 
report Rashba couplings of order $10^{-12}-10^{-11}$\,eV\,m, based on which we take 
$\alpha^{\text max}=\frac{4\gamma}{27\beta}=0.67\times10^{-11}$\,eV\,m.
Assuming that this maximum value is attained for densities $n_0\sim0.1$\,el/u.c.
(for typical experimentsl values of the densities at the interface, see, e.g., Ref. \onlinecite{espci2}), we get 
$\gamma=3.82\times10^{-20}$\,eV$\,$m$^2\,$V$^{-1}$ and $\beta=8.45\times10^{-10}$\,m\;V$^{-1}$. Note 
that this value of $\beta$ corresponds to a maximum field at the interface of 
$\bar{E}=6\times10^{8}$\,V\,m$^{-1}$, comparable to the values estimated in Ref. \onlinecite{espci2}.

A comment is now in order about the choice of the field-dependent RSOC in Eq. (\ref{alpha_E}). Since LXO 
and STO are different materials, inversion symmetry at the interface is broken even in the absence of 
polarity catastrophe reconstruction (i.e., for $n_0=0$). This means that a residual electric field 
besides $E_{pc}$ should always be present, leading to a finite RSOC, $\alpha(n_0=0)$, even for vanishing 
$E_{pc}(n_0=0)$. This additional field, no matter how small, has been shown in Ref. \onlinecite{CPG} to 
always produce phase separation at low enough densities ($n_0\lesssim 10^{12}$ el./u.c.) for the 
linear--in-k RSOC considered there. However, we are interested in the typical regime of real materials 
with densities $n_0\gtrsim 10^{13}$ el./u.c. Therefore, in order to keep the smallest number of model 
parameters, we omit this term, which, in any case, favors the occurrence of phase separation and 
therefore would strengthen our conclusions.

As will become clearer below, the possibility of a phase separation depends only 
marginally on the precise density dependence of the RSOC. The crucial point is that when the 
Fermi energy enters the heavy bands, the coupling and/or its derivative with respect to $n_0$ need to 
exceed a certain value in order for the instability to occur.

%%%   D E N S I T Y    O F    S T A T E S   %%%
\subsection{Chemical Potential}
From the band structure we deduce the density of states (DOS):
\[
g_n(\varepsilon)=\frac{1}{(2\pi)^2}\int\delta\bigl(\varepsilon(k_x,k_y)-\varepsilon\bigr)\,dk_xdk_y,
\]
where the subscript $n$ reminds us that, like the band structure, also the DOS depends on the density.
These density dependencies add some subtleties to the calculation of the chemical potential which we 
will discuss in the following.

We begin by considering the simple example of a constant DOS $g_n(\varepsilon)=g$ and a 
density-dependent band bottom $\varepsilon_0(n)$, in which case the Fermi level ${\varepsilon_F}$ reads 
\begin{equation}\label{E_F}
n=\int_{-\varepsilon_0(n)}^{\varepsilon_F} g\,d\varepsilon \:\Rightarrow \varepsilon_F=\frac{n}{g}-
\varepsilon_0(n),
\end{equation}
and the total energy per unit cell is given by
\[
E=\int_{-\varepsilon_0(n)}^{\varepsilon_F} \varepsilon\,g\,d\varepsilon \:\Rightarrow E=
\frac{n^2}{2g}-n\varepsilon_0(n).
\]
Differentiating once (twice) with respect to the density we obtain the chemical potential 
(inverse compressibility)
\begin{eqnarray*}
\mu&=&\frac{\partial E}{\partial n}=n\Bigl(\frac{1}{g}-\frac{\partial\varepsilon_0}{\partial n}\Bigr)
-\varepsilon_0(n),\nonumber\\
\kappa^{-1}&=&\frac{\partial\mu}{\partial n}=\frac{1}{g}-2\frac{\partial\varepsilon_0}{\partial n}
-n\frac{\partial^2\varepsilon_0}{\partial n^2}.\nonumber
\end{eqnarray*}
The inverse compressibility, instead of being equal to $1/g$ as in the standard case of a constant 
band structure, has two additional contributions coming from the density dependence of the energy. 
Therefore, we find that already in the simplified case of a constant DOS, the density dependence of 
$\varepsilon_0$ introduces new terms to the chemical potential and its derivative.

Next, we address the implications of the density dependence of the DOS. To calculate the 
thermodynamic properties of the system we consider the grand-canonical ensemble defined by the 
potential $\Omega=F-\mu N$, where $F$ is the free energy and $N$ the total particle number of the 
system. In this ensemble the chemical potential is fixed while the particle density 
$n=n_0+\delta n=N/L^2$ is a fluctuating quantity ($L^2$ is the number of lattice sites). 
To be precise, the fluctuating quantity here is $n_0$, while $\delta n$ is univocally fixed by the 
gating $V_g$, which is an external parameter. Therefore, when we calculate the chemical potential via 
the density-dependent grand-canonical potential, the density [and therefore also $\alpha(n_0)$]
is not known a priori. An elegant way to circumvent this problem is the introduction of a 
Lagrange multiplier $\lambda$; we take the energy to be dependent on a real parameter $x$, determine 
the chemical potential $\mu(\lambda,x+\delta n)$, and then chose $\lambda$ to impose $x=n_0$. So, 
we consider the following grand-canonical Hamiltonian: 
\begin{eqnarray*}
{{\cal H}}_{\lambda}(x)&=&  {{\cal H}}(x)-\mu\sum_{k\sigma} {c}_{k\sigma}^{\dagger}
{c}_{k\sigma}+\lambda\sum_{k\sigma}\bigl[{c}_{k\sigma}^{\dagger} {c}_{k\sigma}-(x+\delta n)\bigr]
\nonumber\\
&=&  {{\cal H}}(x)-\mu^*\sum_{k\sigma} {c}_{k\sigma}^{\dagger} {c}_{k\sigma}-\lambda L^2 (x+\delta n),
\nonumber
\end{eqnarray*}
where $c$ and $c^\dagger$ are the usual electron destruction and creation operators,
$ {{\cal H}}(x)$ is the canonical one-particle Rashba Hamiltonian of Eq. (\ref{H_Rashba}) and 
$\mu^*=\mu-\lambda$. The grand-canonical potential density $\omega=\Omega/L^2$ reads
\begin{eqnarray*}
\omega&=&-\frac{T}{ L^2}\text{ln}\Bigl[\,\text{Tr}\Bigl(\text{e}^{- {{\cal H}}_{\lambda}(x)/T}\Bigr)
\,\Bigr]\nonumber\\
&=&-\frac{T}{L^2}\sum_{k\gamma}\text{ln}\Bigl[\,1+\text{e}^{-(\varepsilon_{k\gamma}(x)-\mu^*)/T}\Bigr]
-\lambda (x+\delta n),\nonumber
\end{eqnarray*}
(for simplicity we here take the Boltzmann constant $k_B=1$)
where $\varepsilon_{k\gamma}$ are the eigenvalues of ${{\cal H}(x)}$ given in Eqs. (\ref{disp_rel_iso}) 
and (\ref{disp_rel_aniso}). The electron density is given by
\begin{equation}\label{muT0}
n=-\frac{\partial\omega}{\partial\mu}=-\frac{\partial\omega}{\partial\mu^*}\frac{\partial\mu^*}{\partial\mu}
=\frac{1}{L^2}\sum_{k\gamma}f(\varepsilon_{k\gamma}(x)-\mu^*),
\end{equation}
where $f(\varepsilon)=(\text{e}^{\,\varepsilon/T}+1)^{-1}$ is the Fermi-Dirac function at a temperature
$T$.
The equations for $x$ and $\lambda$ read:
\begin{flalign}
&\frac{\partial\omega}{\partial\lambda}=0\;\Rightarrow n=x+\delta n,\nonumber\\
&\frac{\partial\omega}{\partial x}=0\;\Rightarrow\lambda=\frac{1}{L^2}\sum_{k\gamma}
\frac{\partial\varepsilon_{k\gamma}}{\partial x}f(\varepsilon_{k\gamma}-\mu^*)\nonumber\\
&=\frac{\partial\alpha}{\partial x}\frac{1}{L^2}\sum_{k}\,k\,
\bigl[f(\varepsilon_{k+}-\mu^*)-f(\varepsilon_{k-}-\mu^*)\bigr].\label{lambdaT0}
\end{flalign}
Notice that $\lambda$ depends on both the absolute value of $\alpha$ [via the difference 
$f(\varepsilon_{k+}-\mu^*)-f(\varepsilon_{k-}-\mu^*)$] and its derivative with respect to $x$ 
(i.e., $n_0$). It is also worth noticing that the pseudo-chemical potential $\mu^*$, 
and not the chemical potential $\mu$, coincides with the Fermi energy at $T=0$.

Eventually, we want to know whether for a given experimentally observed value of $n_0$, 
say $n_0^{exp}$, fixed by the chemistry of the sample, the EG is stable or not. 
This amounts to determine whether the chemical potential $\mu$ is an increasing or a 
decreasing function of $n_0$ in the vicinity of $n_0^{exp}$. To accomplish this, we vary $n_0$ over 
the interval $I_0=[0,0.08]$\,el./u.c. for $V_g\geq0$ and $I_0=[|\delta n|,|\delta n|+0.08]$\,el./u.c. 
for $V_g\leq 0$, keeping the gating fixed, and calculate for each density the chemical potential
\[
\mu(n_0;V_g)=\mu^*(n_0;V_g)+\lambda(n_0;V_g),\:\:\:n_0\in I_0,
\]
which yields curves of the form shown in Fig. \ref{mu_n_T0}.
The respective densities $n_{01}$ and $n_{02}$ of the coexisting phases are then determined by the Maxwell
construction
\[
\int_{n_{01}}^{n_{02}}\mu(n_0;V_g)\,dn_0=\bar{\mu}(n_{02}-n_{01}),
\]
where $n_{01}(V_g)$, $n_{02}(V_g)$ and $\bar{\mu}(V_g)$ satisfy the relations 
$\mu(n_{01};V_g)=\mu(n_{02};V_g)=\bar{\mu}(V_g)$. 
In case $n_0^{exp}$ falls into the interval $[n_{01},n_{02}]$, the system is unstable 
and separates into two phases with density $n_1=n_{01}+\delta n$ and $n_2=n_{02}+\delta n$, respectively.
An example of the Maxwell construction is given in the right panel of Fig. \ref{mu_n_T0} for 
$V_g=100$\,V, where $n_{01}\sim0$ and $n_{02}\sim0.04$. The contribution of the Lagrange multiplier
$\lambda$ to the chemical potential and to the compressibility, which was neglected in Ref. 
\onlinecite{CPG}, enhances the tendency to electronic phase separation.

\section{The phase separation at {$T=0$, $B=0$}}\label{Sect_Results}

We commence the study of the phase separation instability with the calculation of the chemical potential 
at zero temperature and zero magnetic field and analyze how it evolves upon varying the gate voltage 
$V_g$. As mentioned above, the electric field due to the charge reconstruction is much larger than the 
one arising from gating and we omit this latter contribution to $\alpha$. Consequently, the only effect 
of increasing (decreasing) $V_g$ is to increase (decrease) the density by $\delta n$ with respect to 
$n_0$, leaving the value of $\alpha(n_0)$ essentially unchanged. Furthermore, we remind that
temperature, magnetic field and gate potential are (the) external parameter determining the 
thermodynamic state of the system, while the density $n_0$ (and its fluctuations) are determined by 
its internal stability. 

\begin{figure}[h]
\begin{center}
\includegraphics[width=1.0\linewidth]{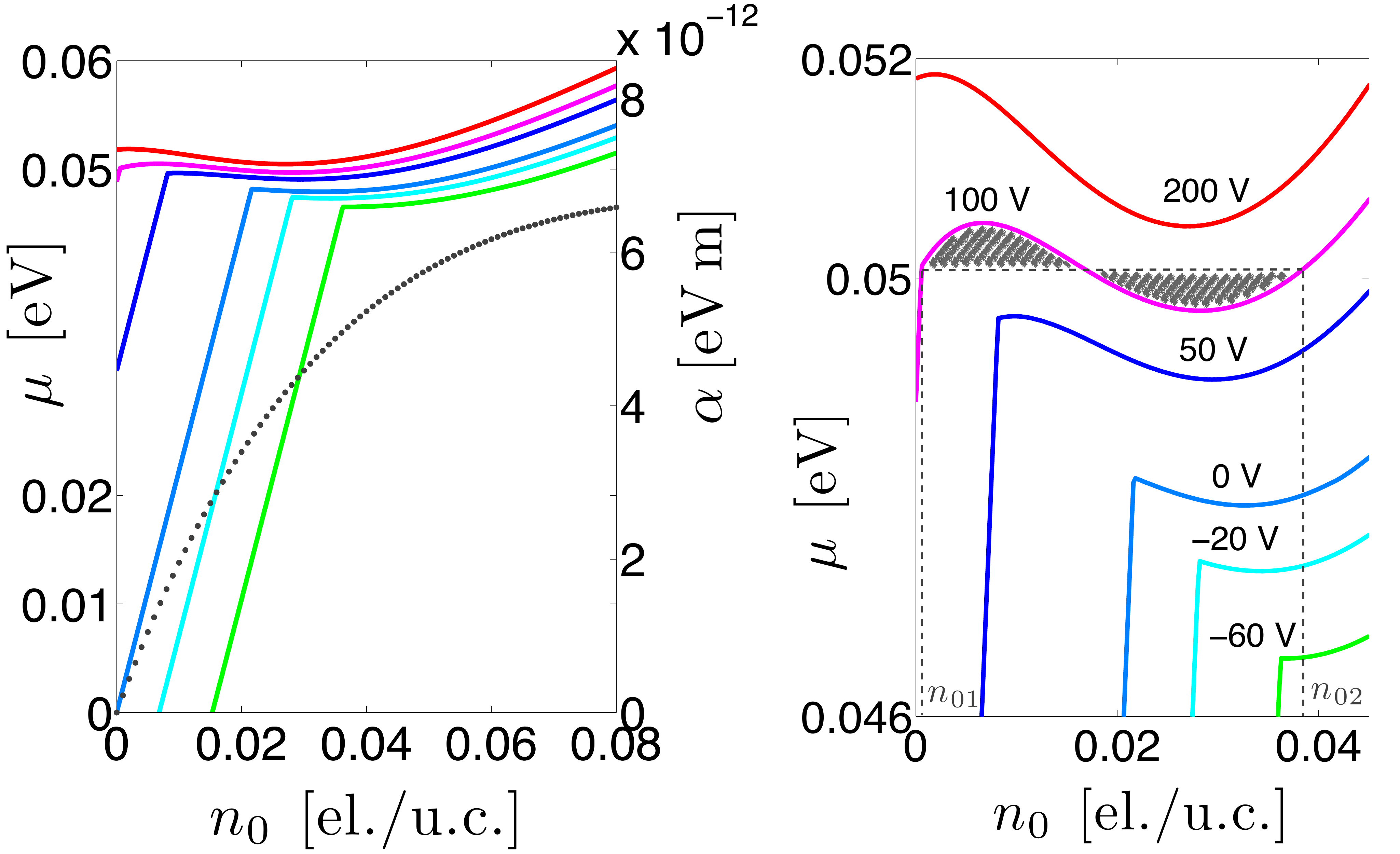}
\caption{(color online) Left panel: $\mu(n_0;V_g)$ for various gate voltages. The green (lowest) curve 
corresponds to the critical curve at a gating $V_g=-60$\,V. The dotted line (and right y-axis) reports 
the Rashba coupling $\alpha(n_0)$ according to Eq. (\ref{alpha_n0}). Right panel: Enlarged view of 
$\mu(n_0;V_g)$ for various gate voltages. The dashed lines and shaded areas report the Maxwell 
construction for $V_g=100$\,V. The values $-60, -20, 0, 50, 100, 200$\,V indicate the gate voltage of 
the corresponding curves.}
\label{mu_n_T0}
\end{center}
\end{figure}

We anticipate here that for the (physically realistic) values of the RSOC considered in this paper, 
no instability occurs as long as only the lower-lying $d_{xy}$ low-DOS band is filled. On the contrary, 
a phase separation instability is {\it possible} at higher electron density, when the heavier 
$d_{xz,yz}$ bands start to be filled. A precondition for this possibility is that the RSOC and 
its derivative with respect to $n_0$ are large enough (see below). Quite remarkably, we find that 
the values of $\alpha(n_0)$ at which the instability occurs are well in the range 
of those inferred from magnetotransport  experiments in LAO/STO\cite{cavigliaPRL,fete} and 
LTO/STO.\cite{espciprivcom} Before considering the occurrence of the instability in a more detailed 
and quantitative way, we address the numerical solutions of $\mu(n_0;V_g)$ shown in Fig. \ref{mu_n_T0} 
and explain them on a qualitative basis.

First of all, one has to bear in mind that, although in the real systems $n_0$ is determined by 
the chemistry of the system, for a given $n_0$ one has to explore a whole interval of $\mu(n_0;V_g)$ 
to establish whether the system stays homogenous at this $n_0$ or whether it is energetically convenient 
to split into separate phases with different electronic reconstructions ($n_{01}$ and $n_{02}$). 
Let us start with a system having, say, $n_0^{exp}=0.015$ at $V_g=0$. Only the  $d_{xy}$ band is filled
and the chemical potential $\mu(0.015;0\,\mbox{V})$ has a positive slope as a function of the electron 
density. Upon increasing $V_g=50$ V other electrons $\delta n >0$ are introduced and the anisotropic 
bands start to be filled, while $\alpha(n_0^{exp}=0.015)$ and $\alpha'(n_0^{exp}=0.015)$ are large enough 
to render the system unstable [negative slope of $\mu(0.015;50\,\mbox{V})$]. The system therefore 
becomes inhomogeneous and splits into regions at densities to be determined by the Maxwell construction 
[see the enlarged view in the right panel of Fig. \ref{mu_n_T0}, where the construction is explicitly 
done for $V_g=100$ V]. If the chemistry were such that $n_0^{exp}=0.08$ in the as-grown system, the  
heavy bands would be filled, but $\alpha(n_0)$ would vary too mildly with $n_0$ and therefore the bands 
would be too rigid to allow for the instability to occur. On the other hand, if the chemistry were such 
that no electrons were transferred at the interface of the as-grown system ($n_0^{exp}=0$), then the 
gated electrons only would fill the bands and one should apply a gate voltage $V_g\approx 100$\,V before 
$\delta n$ is large enough to fill the heavy bands. At this point, upon increasing $V_g$, the system 
becomes unstable and regions with different electronic reconstructions, $n_{01}=0$  and 
$n_{02}\sim 0.04$, are formed. 

Once this rapid survey of the unstable reconstruction has been given, we now move to a more analytic 
understanding of the above phenomenology at zero temperature and zero magnetic field.
The chemical potential for a constant isotropic band of mass $m^*$ is
\begin{equation}\label{musi}
\mu^*(n_0;V_g) = \begin{cases} \frac{(n_0+\delta n)^2}{4g_0^2\varepsilon_0}-\varepsilon_0, & n_0+\delta n
\leq2g_0\varepsilon_0
\vspace{0.1cm}\\ 
\frac{n_0+\delta n}{g_0}-2\varepsilon_0, & n_0+\delta n\geq2g_0\varepsilon_0 \end{cases}
\end{equation}
where $g_0=m^*/\hbar^2\pi$ corresponds to the constant DOS of a 2DEG.
The correction term due to the density dependence of the Rashba coupling reads
\begin{equation}\label{lambdai}
\lambda(n_0;V_g) = \begin{cases} -\frac{\alpha'(n_0)}{6\pi}\Bigl(k_{-2}^3-k_{-1}^3\Bigr), & 
n_0+\delta n\leq2g_0\varepsilon_0\vspace{0.2cm}\\ 
-\frac{\alpha'(n_0)}{6\pi}\Bigl(k_-^3-k_+^3\Bigr), & n_0+\delta n\geq2g_0\varepsilon_0 \end{cases}
\end{equation}
where
\begin{eqnarray*}
k_{-1,2}&=&\frac{m^*}{\hbar^2}\biggl(\alpha\mp\sqrt{\alpha^2+2\hbar^2\mu^*/m^*}\biggr),\nonumber\\
k_{\pm}&=&\frac{m^*}{\hbar^2}\biggl(\pm\alpha+\sqrt{\alpha^2+2\hbar^2\mu^*/m^*}\biggr).\nonumber
\end{eqnarray*}

For the chosen values of the parameters $\beta$ and $\gamma$, the condition 
${\partial \mu(n_0;V_g)}/{\partial n_0}<0$ is never satisfied,\cite{2E0g0} 
and the terms which depend on $\alpha$ and $\alpha'$ [via $\varepsilon_0=\varepsilon_0(\alpha)$] are 
small compared to $(n_0+\delta n)/g_0$. Therefore, the major effect of gating is to translate the 
(quasi) linear curves $\mu(n_0;V_g)$ by $\delta n(V_g)$ as it is shown in Fig. \ref{mu_n_T0}. In terms 
of the total density $n$ a rigid shift by $\delta n$ translates into coinciding 
$\mu(n)$. We note that here we can write $\mu(n)$, because it does not matter whether the electrons come 
from the polarity catastrophe or gating. 

\begin{figure}[h]
\begin{center}
\includegraphics[width=1.0\linewidth]{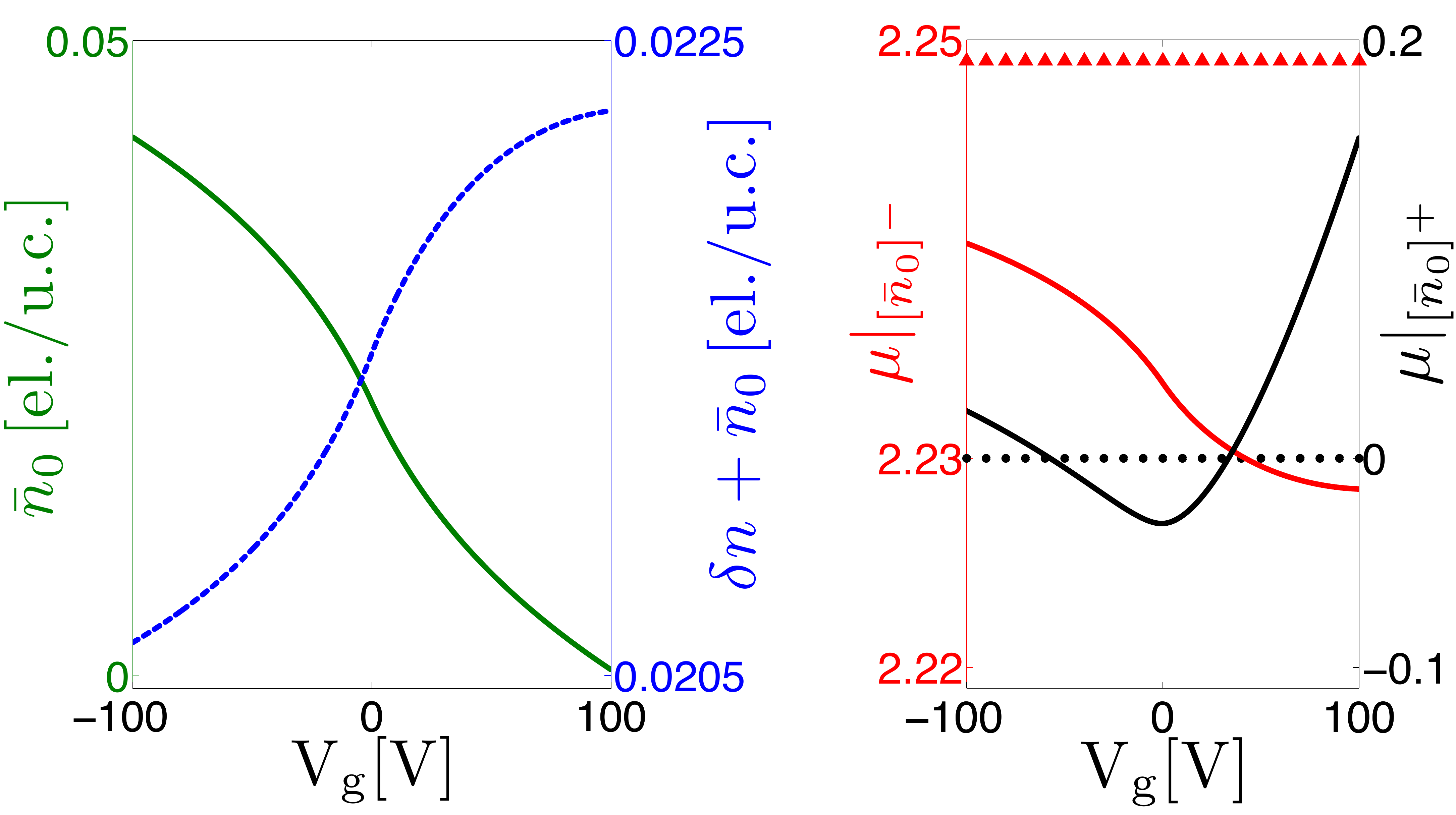}
\caption{(color online) 
Left panel:
The green curve (and left y-axis) reports $\bar{n}_0$ vs. $V_g$ 
obtained from Eq. (\ref{eq:muaband}) which corresponds to the density 
$\bar{n}_0$ at which the anisotropic bands start to be filled.
The same result but with the gating induced density included is shown
by the blue curve (right y-axis). 
Right panel:
The full black curve shows the inverse compressibility 
$\mu'(n_0)$ evaluated as $n_0\to[\bar{n}_0]^+$ (right y-axis). Zeros are indicated
by the black dotted line. The red curve (left y-axis) reports the analogous result but for
$n_0\to[\bar{n}_0]^-$. For comparison, the
red curve with triangles shows the constant inverse compressibility $1/g_0$ of a rigid isotropic band.}
\label{dmu_at_N0bar}
\end{center}
\end{figure}

What happens when the anisotropic bands start to be filled? In general, one expects effects due 
to the RSOC to increase. Indeed, inspecting Fig. \ref{mu_n_T0} one observes a strong flattening of 
the chemical potential when $\mu\sim\Delta$. In order to pinpoint the density at which the anisotropic 
bands begin to be filled, we solve the equation
\begin{equation}\label{eq:muaband}
\mu=\Delta-\varepsilon^{\text{a}}_0,
\end{equation}
where $\varepsilon^{\text{a}}_0=m_h\alpha^2/(2\hbar^2)$.  
The solutions $\bar{n}_0$ and $\bar{n}=\bar{n}_0+\delta n$ are reported in the left plot of Fig. 
\ref{dmu_at_N0bar} and correspond, as expected, to the densities at which the chemical potential shown in 
Fig. \ref{mu_n_T0} changes abruptly. 
For $V_g\gtrsim100$\,V we find $\bar{n}_0\sim0$, meaning that $\delta n$ fills up the isotropic 
band to the bottom of the anisotropic bands.
%%%%%%%%%%%%%%%%%%%%%%%%%%%%%%%%%%%%%%%%%%%%%

In general, the equations for $\mu>\Delta-\varepsilon^{\text{a}}_0$ do not allow for a simple solution 
due to the absence of rotational symmetry. If, however, we are only interested in the region 
$n_0\gtrsim\bar{n}_0$, we can expand $\varepsilon^{\text{a}}_-(k)$ around its four minima 
${\bf{k}}^{xz}_0=(\pm\frac{m_h\alpha}{\hbar^2},0)$ and 
${\bf{k}}^{yz}_0=(0,\pm\frac{m_h\alpha}{\hbar^2})$. Due to invariance of the Hamiltonian under a 
rotation of $\pi/2$, it is sufficient to consider one minimum only (and multiply by $4$ at the end). 
Taking, for example, ${\bf{k}}_0=(\frac{m_h\alpha}{\hbar^2},0)$, we obtain
\[
\varepsilon^{\text{a}}_-(k_x,k_y)\approx\Delta-\varepsilon^{\text{a}}_0+\frac{\hbar^2}{2m_x}
\Bigl(k_x-\frac{m_h\alpha}{\hbar^2}\Bigr)^2+\frac{\hbar^2k_y^2}{2M}
\]
where $M=m_xm_y/(m_x-m_y)$. 

Based on this quadratic dispersion relation we derive the following quantities, valid for 
${\bf{k}}\sim{\bf{k}}_0$:
\begin{flalign*}
&g^{\text{a}}_0=\frac{\sqrt{m_xM}}{\pi\hbar^2},\nonumber\\
&\mu^*(n_0;V_g) =\frac{1}{g^{\text{i}}_0+g^{\text{a}}_0}\Bigl[(n_0+\delta n)
-2g_0^{\text{i}}\varepsilon_0^{\text{i}}
+(\Delta-\varepsilon^{\text{a}}_0)g^{\text{a}}_0\Bigr],\nonumber\\
&\lambda(n_0;V_g)=-\alpha'\Biggl[\frac{\alpha\sqrt{m_xM}}{\pi\hbar^2}\biggl( \frac{2m_x}{\hbar^2} 
\bigl(\mu^*-\Delta+\varepsilon^{\text{a}}_0\bigr) \biggr)\nonumber\\
&\quad\quad+\frac{\hbar^2}{8\pi\alpha m_x}\biggl(\frac{M}{m_x}\biggr)^{3/2}\biggl(\frac{2m_x}{\hbar^2} 
\bigl(\mu^*-\Delta+\varepsilon^{\text{a}}_0\bigr)\biggr)^2\biggr].\nonumber
\end{flalign*}

In addition, we calculate the chemical potential for densities $n_0\gtrsim\bar{n}_0$,
\begin{eqnarray*}
n_0+\delta n&=&\int_{-\varepsilon_0^\text{i}}^{\mu^*}g^{\text{i}}_0(\varepsilon)\,
d\varepsilon+\int_{\Delta-\varepsilon^\text{a}_0}^{\mu^*}g^{\text{a}}_0(\varepsilon)\,
d\varepsilon\nonumber\\
&=&\bar{n}+\frac{m_l}{\hbar^2\pi}(\mu-\lambda)+\frac{2\sqrt{Mm_h}}{\hbar^2\pi}
(\mu-\lambda+\varepsilon^{\text{a}}_0)\nonumber
\end{eqnarray*}
from which we deduce the value of the inverse compressibility,
\begin{equation}\label{dmu}
\frac{\partial\mu}{\partial n_0}=\frac{1}{X}\Biggl[1+X\frac{\partial\lambda}{\partial n_0}-
\frac{2\sqrt{Mm_h}}{\hbar^2\pi}\frac{\partial\varepsilon^{\text{a}}_0}{\partial n_0}\Biggr],
\end{equation}
where $X=(m_l+2\sqrt{Mm_h}\,)/(\hbar^2\pi)$. 
Evaluating Eq. (\ref{dmu}) at $\bar{n}_0$ we obtain the (inverse) compressibility when 
the anisotropic bands start to be filled, which is shown in the right plot of Fig. \ref{dmu_at_N0bar}.
We find that at large negative voltage $V_g\sim-100$\,V, $\mu'(\bar{n}_0)>0$, i.e., the EG is stable. 
Increasing the voltage we arrive at a critical value $V_{gc}\approx-60$\,V where $\mu'(\bar{n}_0)$ 
changes sign, which means that the EG becomes locally unstable.
Increasing further the gate voltage, it becomes stable again at $\bar{n}_0$, 
and the phase separation occurs at higher density (see also Fig. \ref{mu_n_T0}).
For comparison, we plot the derivative at at the top of the isotropic band. One observes 
that the derivative is close to $1/g_0^\text{i}$ which corresponds to its value for a constant band.
This confirms that the contribution from $\alpha$ and $\alpha'$ are indeed small in the isotropic band.

\begin{figure}[h]
\begin{center}
\includegraphics[width=1.0\linewidth]{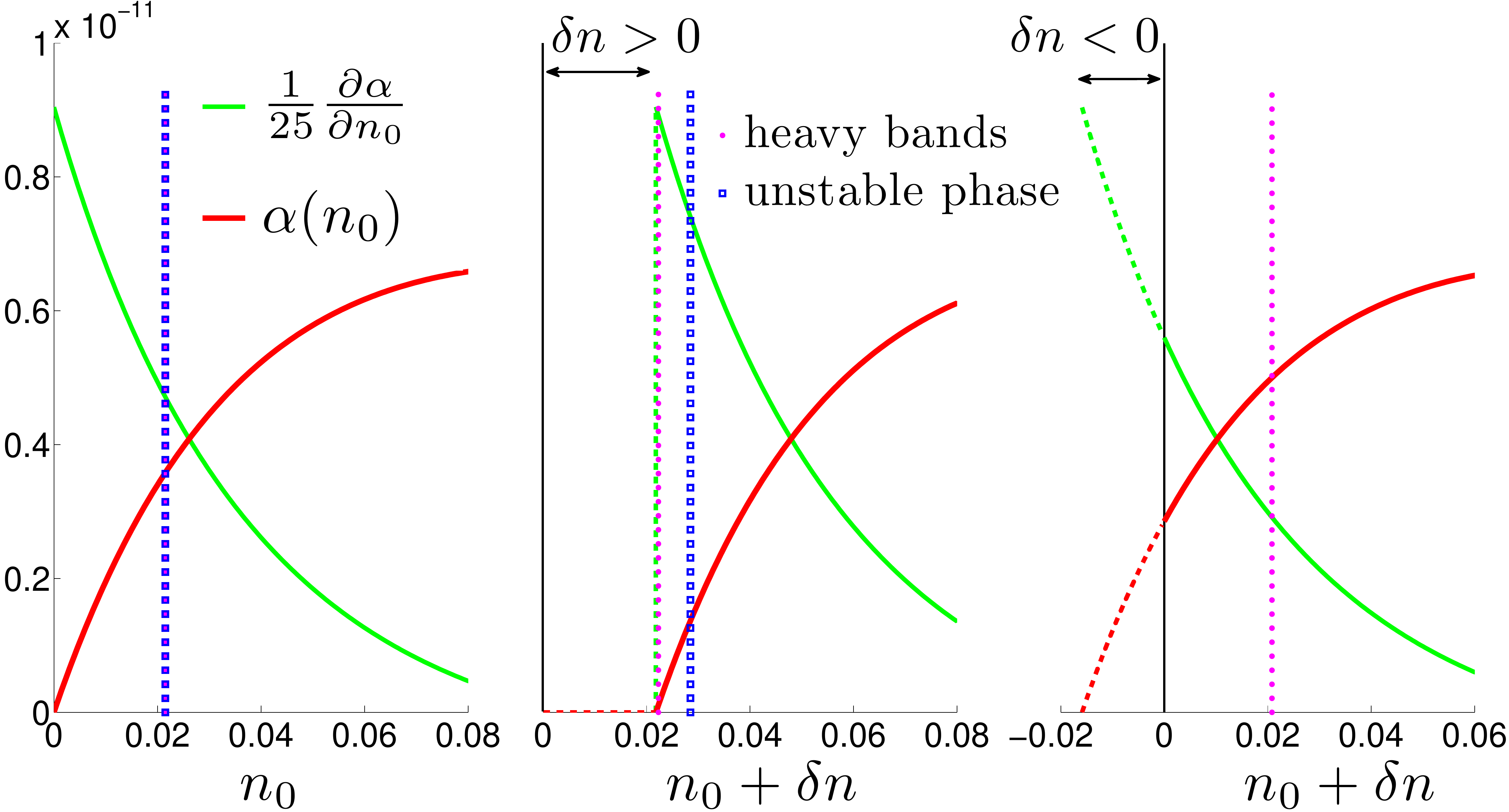}
\caption{(color online) Rashba coupling $\alpha$ (red) and its derivative $\alpha'$ (green) 
as a function of the total density $n=n_0+\delta n$ at a gating $0$\,V (left), $100$\,V (middle), 
$-60$\,V (right). The dots and square markers correspond to the densities $n$ at which the 
anisotropic bands start to be filled and at which the region with $\partial\mu/\partial n<0$ is reached, 
respectively.}
\label{plot_alpha}
\end{center}
\end{figure}

To summarize the above findings we plot in Fig.  \ref{plot_alpha} the density dependence of 
$\alpha$ (red curve) and $\alpha'$ (green curve) for $V_g=0$\,V, $V_g=100$\,V and $V_g=-60$\,V (from 
left to right).
We also include $\bar{n}_0$ (dots) and the density $\tilde{n}_0$ at which the chemical potential attains 
its local maximum (square markers). At zero gating, the blue and purple lines coincide, i.e., the EG 
becomes unstable as soon as the anisotropic bands start to be filled. Lowering the gating down to 
$V_g\sim-60$\,V the phase separation vanishes. Fig. \ref{plot_alpha} reveals that 
$\alpha[\bar{n}_0(V_g=-60$\,V$)]>\alpha[\bar{n}_0(V_g=0)]$ and 
$\alpha'[\bar{n}_0(V_g=-60$\,V$)]<\alpha'[\bar{n}_0(V_g=0)]$ thereby illustrating the 
importance of a sizable derivative of the Rashba coupling with respect to density (i.e., a sufficiently 
non-rigid band) in order to drive the system unstable. Increasing the gating up to $V_g\sim100$\,V the 
EG remains stable upon initially filling the anisotropic bands and it is 
only after further augmenting the density that the instability occurs. The explication is identical to 
the previously discussed case, with the roles of $\alpha$ and $\alpha'$ interchanged. According to Fig.  
\ref{plot_alpha}, $\alpha'[\bar{n}_0(V_g=100$\,V$)]>\alpha'[\bar{n}_0(V_g=0)]$ and 
$\alpha[\bar{n}_0(V_g=100$\,V$)]<\alpha[\bar{n}_0(V_g=0)]$ with $\alpha[\bar{n}_0(V_g=100$\,V$)]$ 
yet to small to cause the instability so that the density has to be augmented by going to larger 
$V_g$'s (or the chemistry should have been different, providing a larger $n_0=\tilde{n}_0>\bar{n}_0$) 
to attain a negative compressibility.
%%%%%%%%%%%%%%%%%%%%%%%%%%%%%%%%%%%%%%%%%%%%%%%

\section{Finite temperature effects}   
The equations relating the particle density to the chemical potential $\mu$ and to the
Lagrange multiplier $\lambda$ at finite temperature maintain the same form as at $T=0$;
only the Fermi-Dirac distribution ceases to be step-like and acquires a temperature dependence.
This dependence leads to a mixing of states with an energy $|\varepsilon-\mu^*|\sim k_BT$ when 
calculating the integrals in Eqs. (\ref{muT0}) and (\ref{lambdaT0}).
As a consequence, the transition from the isotropic to the anisotropic bands smoothens.
\begin{figure}[h]
\begin{center}
\includegraphics[width=1.0\linewidth]{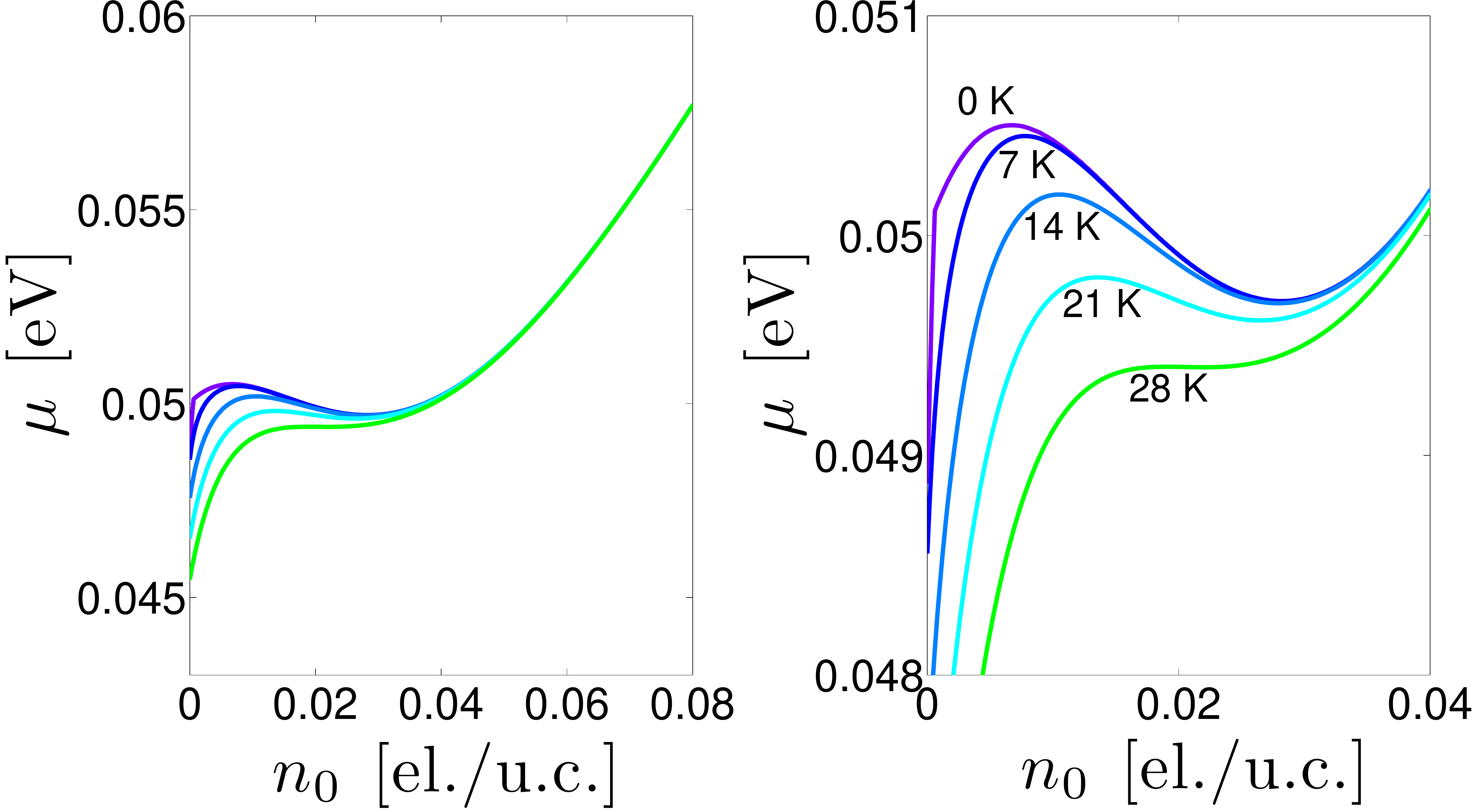}
\caption{(color online) Left panel: Thermal evolution of $\mu(n_0;V_g)$ for $V_g=100$\,V. 
The green (lowest) curve corresponds to the 
critical curve at $T_c\approx28$\,K. Right panel: Zoom of the left panel to 
give an enlarged view of the unstable region.}
\label{mu_vs_n_V100_Va}
\end{center}
\end{figure}
More importantly, the phase separation, if present at $T=0$\,K, vanishes at a temperature $T_c$ of the 
order of $\delta\mu/k_B$ equal to the difference between the two spinodal points divided by the 
Boltzmann constant. This means that the energy associated to the thermal excitations is comparable to 
the energy the EG gained when it phase separated. In Fig. \ref{mu_vs_n_V100_Va} we report the thermal 
evolution of $\mu(n_0;V_g)$ for $V_g=100$\,V and find $\delta\mu\approx0.8$\,meV and
$k_BT_c\approx2.4$\,meV.

\begin{figure}[h]
\begin{center}
\includegraphics[width=1.0\linewidth]{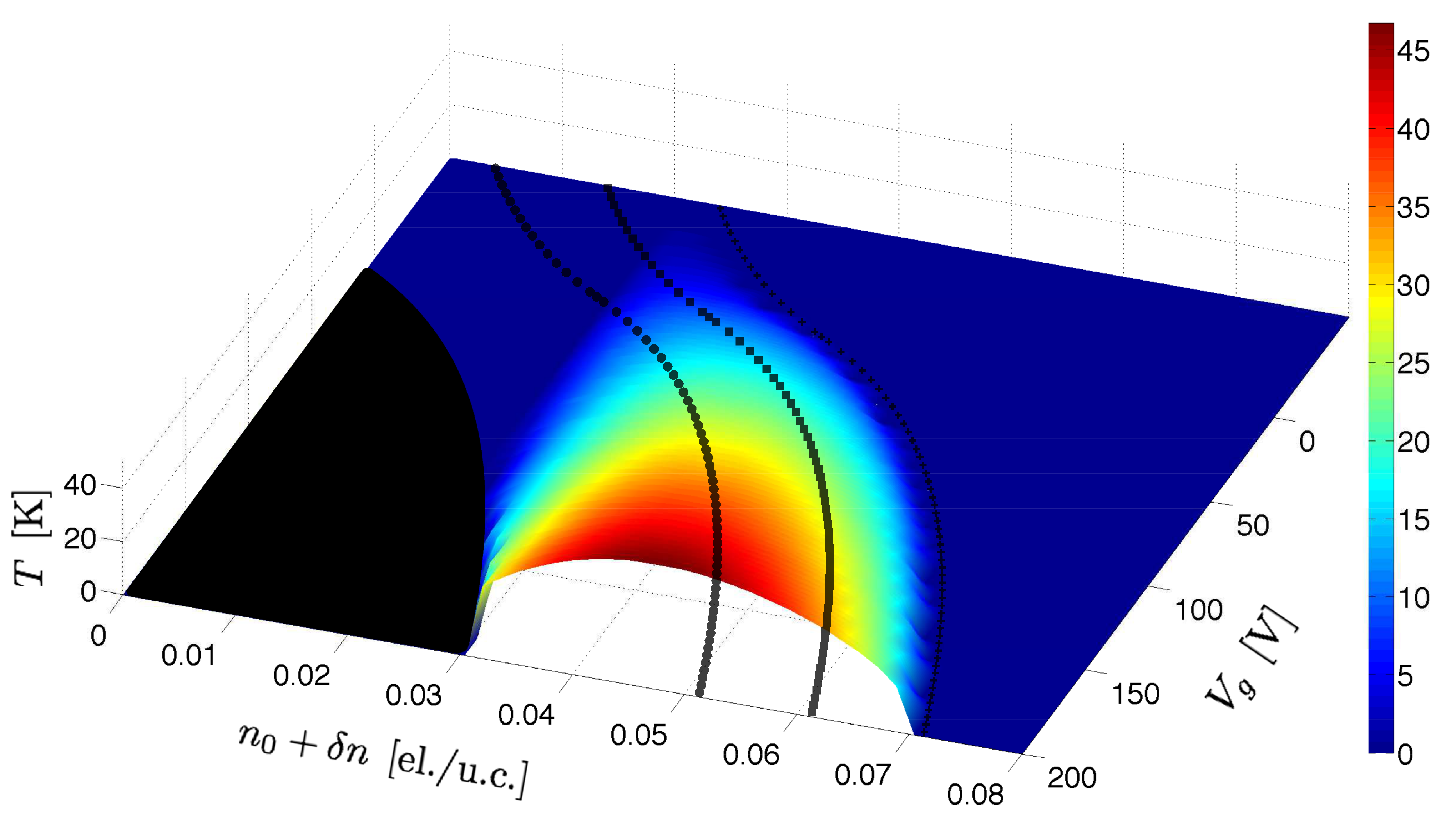}
\caption{(color online) Phase diagram: Electronic phase separation as a function of temperature. 
The markers correspond to 
(from left to right) $n_0=0.02$, $n_0=0.03$ and $n_0=0.04$\,el/u.c.. In this $(n,V_g,T)$ 
plot the system is unstable below the dome-shaped surface and the black shaded area corresponds to the 
physically inaccessible region $n_0<0$.}
\label{Ph_diag_T}
\end{center}
\end{figure}

In Fig. \ref{Ph_diag_T} we report the evolution of the phase separated densities $n_1$ and $n_2$ 
in temperature for $V_g\in[-60,200]$\,V. The obtained phase diagram has a dome-like shape separating 
the stable from the unstable phase. The black shaded area corresponds to the physically inaccessible 
region $n_0<0$ and the markers to $n(V_g)$ at $n_0$ fixed, which represents the behavior of the density 
in gating and temperature for an actual sample (where $n_0$ is fixed by its specific chemistry). We 
find that for realistic values of $n_0\sim[0.01,0.05]$\,el/u.c., the EG is unstable in a large range 
of gating and (at $T=0$\,K) exits the phase separation dome near the quantum critical point at 
$V_g\approx-60$\,V.\cite{bianconi}
This is an interesting occurrence because it implies that near this gating value the system is 
characterized by strong nearly-critical quantum (i.e., intrinsically dynamical) fluctuations of 
its reconstructed structure. This entails strong density fluctuations which at zero temperature 
display critical dynamical behaviors with large dynamical critical exponents $z=3$ (for clean 
nearly ballistic electrons) or $z=4$ (for diffusive electrons). 
Of course, these critical behaviors rest on the assumption that these large scale density fluctuations
are neutral at long distance because they are charge compensated by other mechanisms (like, e.g.,
variations of the Ti, Al valency), which should still be able to follow the slow critical 
electronic dynamics. The possible consequences of slow 
nearly critical density fluctuations on the superconducting critical behavior of these oxide 
interfaces at low gating/densities are presently under investigation.
%%%%%%%%%%%%%%%%%

\section{Effects of a finite magnetic field parallel to the interface, $B\parallel xy$}\label{magn-field}
In this section we analyze the effects of a magnetic field on the electronic phase separation at zero 
temperature. We limit our considerations to the simpler case of an \emph{in-plane} magnetic field,
where the magnetic field is parallel to the interface and couples to the electron spin only, 
leaving the orbital part unaltered. The dispersion relations read:
\begin{flalign}
\:\:&\varepsilon^\text{i}_{\pm}(k,\phi)=\frac{\hbar^2k^2}{2m}\nonumber\\
&\qquad\quad\pm\sqrt{\alpha^2k^2+2\alpha k\mu_BB\sin(\phi)+\mu_B^2B^2},
\label{EBi}\\
\:\:&\varepsilon^\text{a}_{\pm}(k_x,k_y,\phi)=\Delta+\frac{\hbar^2k_x^2}{2m_x}+\frac{\hbar^2k_y^2}{2m_y}
\nonumber\\
&\qquad\qquad\quad\pm\sqrt{\alpha^2k^2+2\alpha k\mu_BB\sin(\phi)+\mu_B^2B^2},\label{EBa}
\end{flalign}
where $\mu_B$ is the Bohr magneton, $B=|{\bf{B}}|$ is the strength of the magnetic field, and 
$\phi$ denotes the angle between $\bf{k}$ and $\bf{B}$. As the equations suggest, the presence of a 
magnetic field leads to a rich band structure. Indeed, the evolution of the band structure differs from 
the previous case where the minima were fixed at ${\bf k}_0^\text{i,a}$. Here, the positions and the 
number of minima depend on the values of $m_x$, $m_y$, $\alpha$, $B$ and $\phi$. We will not go into 
details of the band structure, but focus instead on the general features important to understand 
the effects of the magnetic field. 

\begin{figure}[h]
\begin{center}
\includegraphics[width=0.95\linewidth]{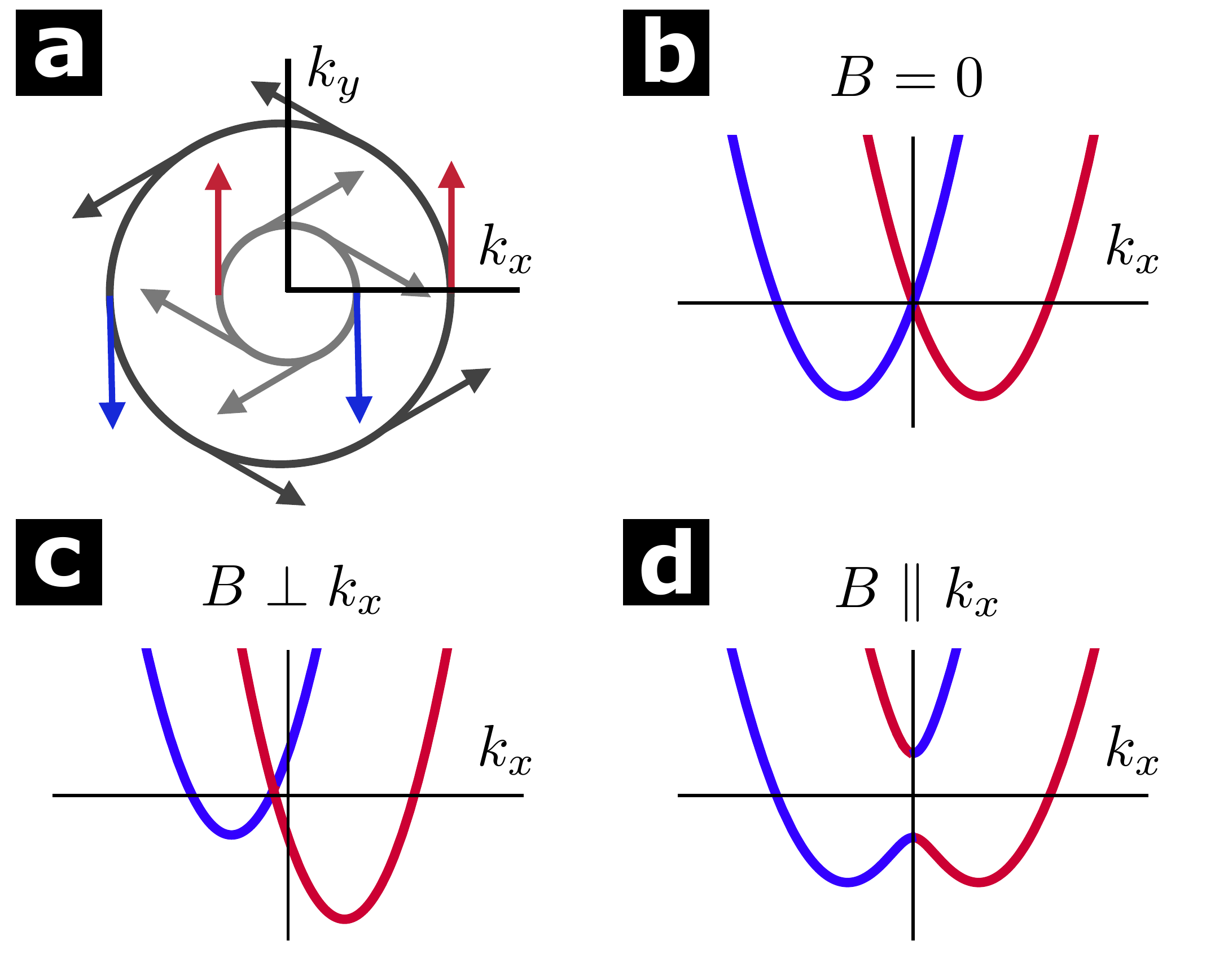}
\caption{(color online) Panels $a,b$: Spinor structure on the Fermi surface and band structure of 
the RSOC system for $B=0$. Panels $c,d$: Resulting band structure for an applied magnetic field 
along the y- and x-direction.}
\label{BandStructure_Bfield}
\end{center}
\end{figure}

The Fermi surface and the band structure along $k_x$ of the isotropic band in the absence of 
a magnetic field are displayed in Fig.  \ref{BandStructure_Bfield} (panels $a$ and $b$). From the 
spin structure of the Fermi surface it is readily seen that one has a competition between the 
chiral structure of the RSOC and the magnetic field which tends to polarize the spins along its 
direction. As a result the application of a magnetic field $\bf{B}$ causes a band shift for states 
$\bf{k}$ perpendicular to $\bf{B}$ (panel $c$) and a Zeeman splitting for states parallel 
to the field (panel $d$).

\begin{figure}[h]
\begin{center}
\includegraphics[width=1.0\linewidth]{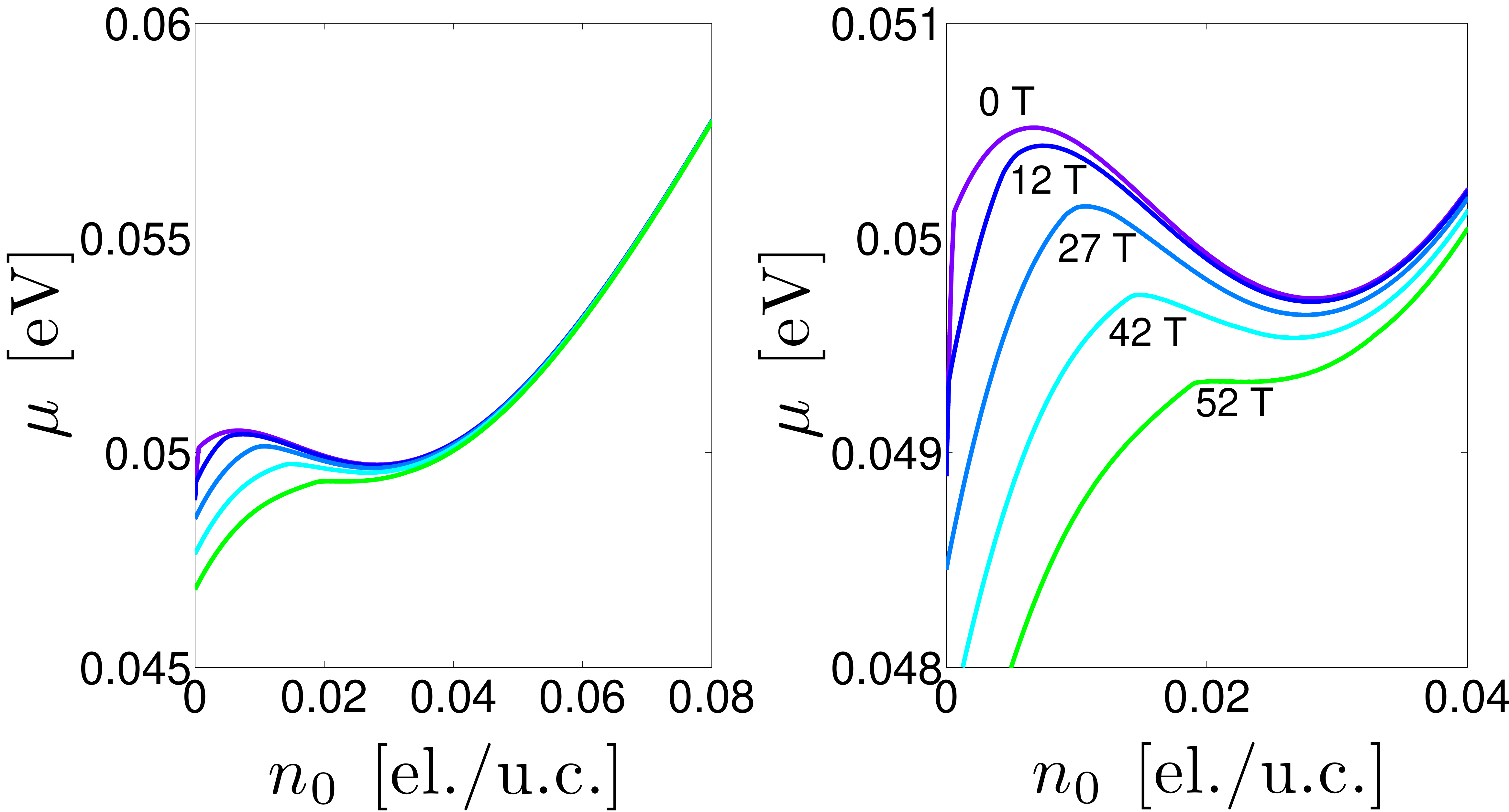}
\caption{(color online) Left panel: Evolution of $\mu(n_0;V_g)$ for $V_g=100$\,V with the magnetic 
field. The green (lowest) curve corresponds to the critical curve at $B_c\approx57$\,T. Right panel: 
Zoom of the left panel to give an enlarged view of the unstable region.}
\label{mu_vs_n_V100_Vb}
\end{center}
\end{figure}

According to Eqs. (\ref{EBi}) and (\ref{EBa}), the phase separation, if it occurs for $B=0$, vanishes 
for magnetic fields of the order $B_c\sim\alpha k_0/\mu_B$. Physically, this means that the magnetic 
field polarizes the spins so strongly, that the effect of the RSOC, which is central for the 
phase separation, is suppressed. As an example, the evolution of the chemical potential with 
increasing magnetic field at gating $100$\,V  is given in Fig. \ref{mu_vs_n_V100_Vb}, where 
$\delta\mu\approx0.8$\,meV and $\mu_BB_c\approx3.3$\,meV.

\begin{figure}[h]
\begin{center}
\includegraphics[width=1.0\linewidth]{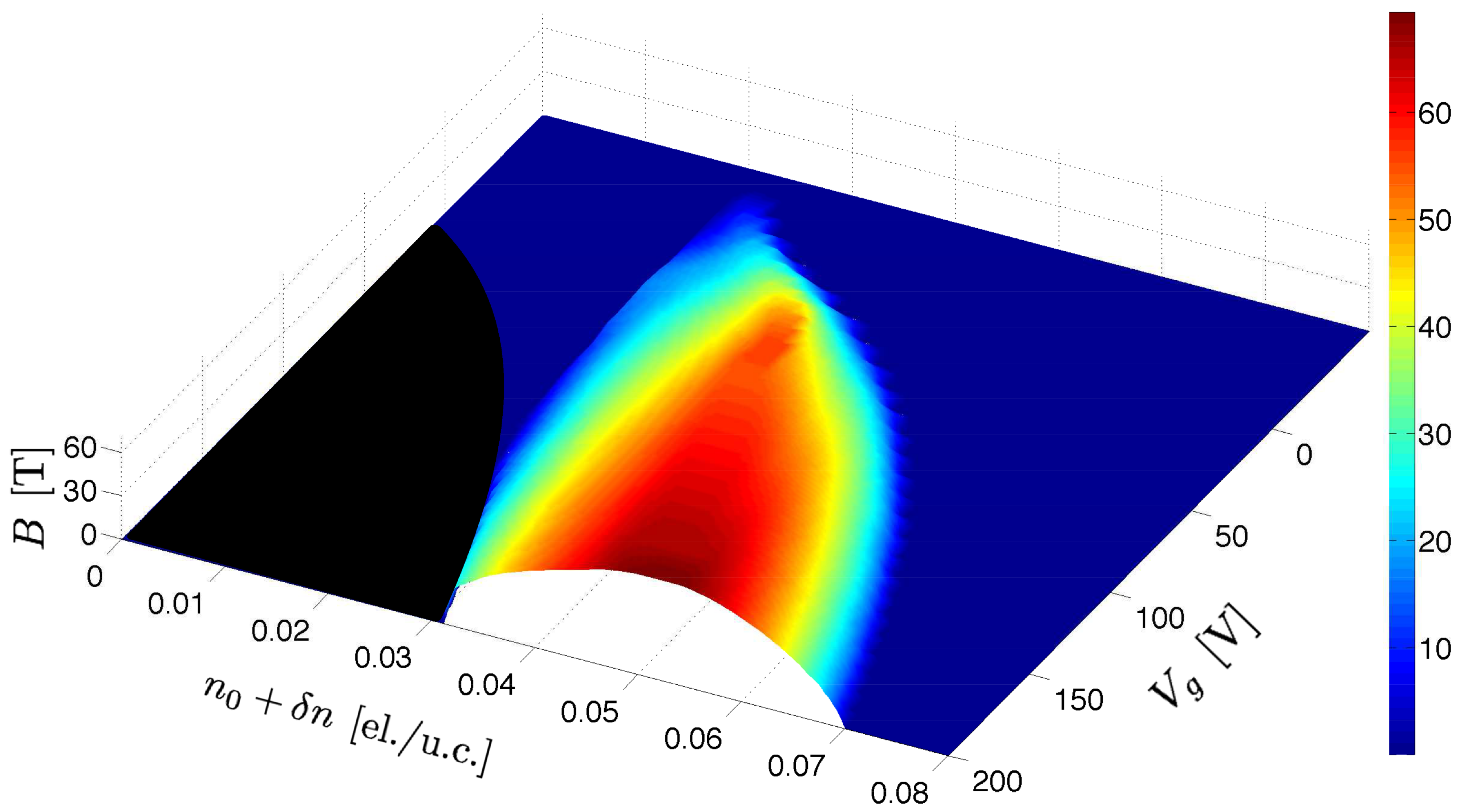}
\caption{(color online) Phase diagram similar to Fig. \ref{Ph_diag_T}
but with the magnetic field along the x-axis and with a temperature $T=0$.}\label{Ph_diag_B}
\end{center}
\end{figure}

In Fig. \ref{Ph_diag_B} we report the evolution of the phase separated densities $n_1$ and $n_2$ in 
magnetic field for $V_g\in[-60,200]$\,V. Similar to the evolution in temperature, we obtain a phase 
diagram with a dome-like shape separating the stable from the unstable phase.
Being at zero temperature, the locus of the critical fields at which phase separation disappears,
as a function of the gate voltage $V_g$, defines a \emph{line} of quantum critical points.
The black shaded area corresponds to the physically inaccessible region $n_0<0$. 
Again we find regions in the phase diagram where strong density fluctuations are expected. Also in 
this case, slow critical quantum fluctuations with $z=3$ or $z=4$ are expected provided the long-range
Coulomb interactions are ineffective and do not spoil the whole phase separation structure.

%%%%%%%%%%%%%%%%%ROLE OF COULOMB
\section{Effects of Coulombic interactions}\label{Sect_Coulomb}   
\begin{figure}[h]
\begin{center}
\includegraphics[scale=0.25]{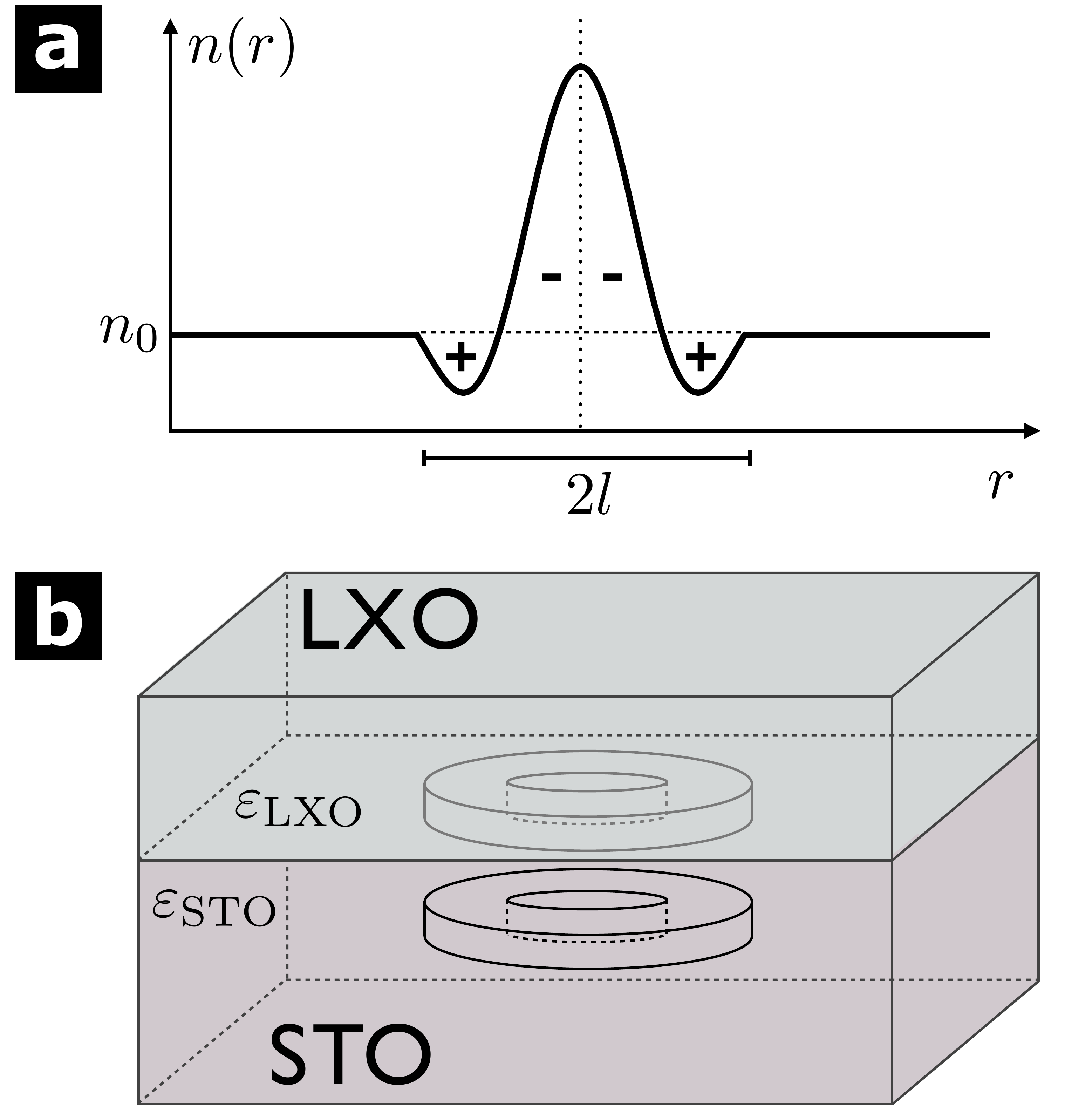}
\caption{(color online) 
(a) Cross section of the excess positive and negative charge density with respect to the 
uniform surface charge density $n_0$ that the polarity catastrophe forms at the LXO/STO 
interface. (b) Schematic view of the excess charge disk+ring residing on the STO side and 
its image residing on the LXO side, used to compute the electrostatic potential. For the 
sake of a clearer display, we draw the disk+ring with diameter $2l$ and a finite thickness, 
while in the calculation this thickness was set to zero.}
\label{fig0}
\end{center}
\end{figure}

The phase separation analyzed above involves the segregation of charged carriers, entailing an 
energetic cost due to Coulombic interactions. In a first step one can assume that, while electrons 
acquire a negative compressibility and tend to segregate, the background of positive countercharges 
stays infinitely rigid. This causes a local charge imbalance which eventually stops the charge 
segregation at finite length scales as soon as the electrostatic cost overcomes the energy gain due 
to phase separation. This so-called frustrated phase separation was extensively investigated in the 
context of cuprates as a possible mechanism of charge inhomogeneous (stripe) 
states.\cite{emerykivelson,FPS} The issue then becomes whether the inhomogeneous state occurs on 
large scales comparable to the experimentally observed nano/mesoscopic disorder (for instance in Ref. 
\onlinecite{doublecrit} a typical size of inhomogeneous regions of order 50-100 nm was estimated) or 
on shorter scales. In the latter case, it might even happen that the scales are so short that 
inhomogeneities simply do not occur and phase separation is spoiled. To complement our analysis of the 
phase separation mechanism, we therefore estimate, as a function of size, the electrostatic cost
of a disk with excess electron charge $-\delta Q$ embedded in the STO matrix, but close to the LXO/STO 
interface. A positively charged ring of excess charge $\delta Q$, arising from a depletion of the 
electron density, is also considered, giving rise to a ``disk+ring'' with an average charge density 
equal to its surrounding. [see Fig.  \ref{fig0}(a)]. For the sake of concreteness, we take for the 
excess electron density the form $n(r)=(A/r)\,\text{sin}(2\pi r/l)$, where $r\in[0,l/2]$ corresponds 
to the disk and $r\in[l/2,l]$ corresponds to the ring.
The electrostatic potential generated by this charge distribution is calculated using the
standard  image-charge method considering the limiting case $\varepsilon_{STO}\gg \varepsilon_{LXO}$
[the dielectric constant of STO is strongly field-dependent, see Eq. (\ref{epsSTO}), 
but always at least one order of magnitude larger than the dielectric constant of the nearby LXO]. 
The charges in STO are taken at a typical distance $d\sim 0-5$ nm from the interface (this distance 
turns out to be much smaller than the disk+ring diameter $2l$). The cost of this inhomogeneous 
charge distribution is then compared to the energy gain from phase separation, as determined from our 
$\mu$ vs. $n$ curves obtained above. We find $l\approx a\,\varepsilon_{\text{STO}}/30$, where $a$ is the 
linear size of the STO unit cell. Taking for $\varepsilon_{\text{STO}}$ the geometric average of the 
minimal ($\sim300$) and maximal ($\sim25000$) value, we obtain $l\sim35-40$\,nm. This means that even 
in the worst case of an infinitely rigid background, sizable nanoscopic inhomogeneities
are produced by phase separation despite the Coulombic cost of charge segregation.

\begin{figure}[h]
\begin{center}
\includegraphics[width=1\linewidth]{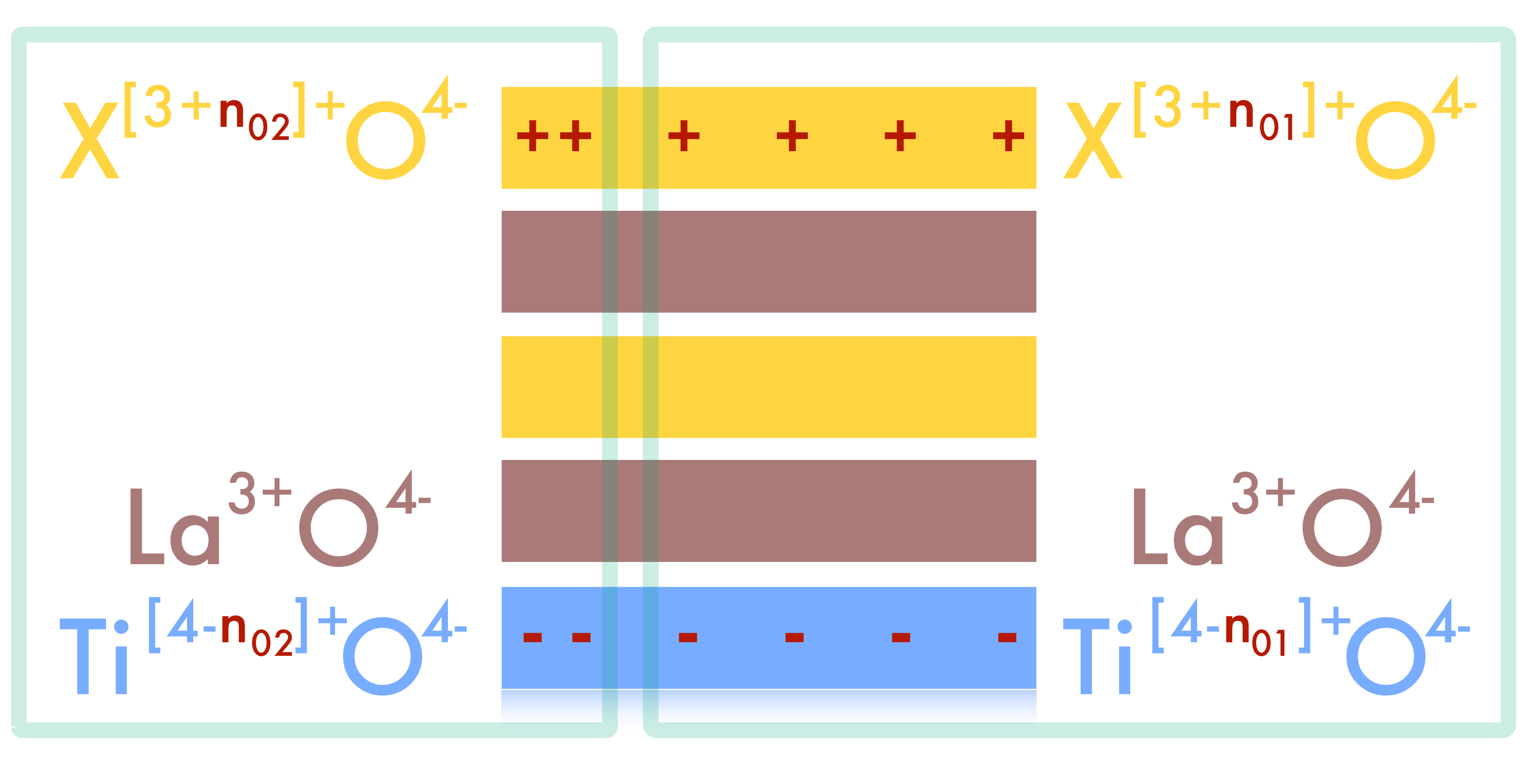}
\caption{(color online) 
Illustration of the phase separation of the LXO/STO interface into two phases with densities $n_{01}$ 
and $n_{02}$, respectively. The positive charges are distributed according to the 
negative ones, thereby rendering the system neutral on length-scales greater than the LXO layer-thickness.
The charge modulation in the top (yellow) XO$_2$ layer does not correspond to a redistribution of X 
ions but to a redistribution of the valency of X.}
\label{PolarCat_withoutCoulomb}
\end{center}
\end{figure}

The worst-case situation sets a lower bound to the size of inhomogeneities. Assuming, as it seems to 
be more realistic, that  the charge distribution in the LXO/STO heterostructure is not rigid,
the possibility  arises that an inhomogeneous distribution of electrons at the interface is (at 
least partially) balanced by a corresponding redistribution of opposite countercharges at the top of 
(and possibly inside) the LXO film, leading to an increase of the size of the  inhomogeneous regions. 
We stress that these adjustments are {\it not} necessarily due to any displacement of ionic positions, 
but may simply arise from small variations of the X$=$Al,Ti ion valency [see Fig. \ref{PolarCat_withoutCoulomb}].
Given the smallness of $n_0$, we can assume that the energetic cost of this even smaller modulation 
of the valency of X is not exceedingly large. Then, the thickness of the LXO film being small, 
the negative charges at the interface and the positive countercharges at the top of the LXO film can 
be considered as a neutral system over distances larger than the thickness of the LXO layer. 

Of course one should worry about the fact that inside the LXO layer the inhomogeneous distribution 
of charges produces an inhomogeneous electric field distribution like in joined capacitors with 
different surface charges on their plates (see Fig. \ref{PolarCat_withoutCoulomb}). This will generate 
an increase of electrostatic energy with respect to the homogeneous case, likely overcoming the 
phase separation gain. However, one should bear in mind that i) the charge reconstruction precisely 
occurs {\it to avoid} the presence of an average electric field in the interstitial polar planes of LXO 
and ii) this reconstruction is only partial because the interface charge, even taking into account 
possible trapping of charge carriers, never coincides with the nominally required 
$n_0=0.5$ el./u.c. (the density of charge carriers is roughly ten times smaller). 
The real amount $n_0$ of charges transferred from the LXO top layer (or bulk) to the interface region 
is unknown. Most of these charges are either trapped at STO impurity states or simply they are not 
transferred because 0.5 is the nominal limit value expected when a (macroscopically) thick stack of 
LXO planes is deposited on the STO substrate, while for smaller thickness (1-10 layers) the systems 
finds a compromise between having a smaller but finite average electric field in LXO and
the energy cost involved in transferring charge to the interface. In this complex situation 
it is really awkward to estimate, e.g., the different energy cost for transferring to the interface, 
say, 0.4 el/u.c. uniformly or creating inhomogeneous regions with 0.38 and 0.42 el/u.c. This estimate 
would involve: a) DFT calculations keeping track of the rigidity of small valence fluctuations of Ti 
ions; b) the screening associated with tiny lattice deformations of the LXO planes; c) the
possibility for a slightly non uniform redistribution of the charge in the intermediate ionic 
planes: the nominal polarity catastrophe distribution (0.5) ($-$1) (+1) ... ($-$1) (+1) ($-$0.5) could 
well become, say, (0.6) ($-$0.9) (+0.8) ... ($-$0.9) (+0.8) ($-$0.4). Moreover, all these
effects related to the polarity catastrophe could well coexist and cooperate with a not fully
uniform distribution of the oxygen vacancies.  An analysis of all these issues is far 
beyond the scope of our paper and an analysis by means, e.g., of DFT calculations, would be
made difficult by the smallness of the charge imbalances and of the related energies involved 
(a few meV) so that a first-principle analysis would be very delicate.

It is however conceivable that small redistributions of the charge in the top 
layer and in the internal polar planes of LXO balance the excess (or lack) of electrons in the 2DEG 
without creating substantial changes in the electric field inside the LXO.  
In this respect, we point out that the worst-case scenario of 
a rigid background already gives rise to large enough inhomogeneities to justify our analysis of 
phase separation. {\it Additional} mechanisms related to the charge flexibility of these systems can only 
extend the effects of phase separation in generating the experimentally  observed inhomogeneous
interface state.

%%%%%%%%%%%%%%%CONCLUSIONS 
\section{Conclusions}\label{Sect_Conclusion}   
Based on a simple and general model we demonstrated that, as soon as the Rashba coupling depends on 
the density, the possibility for an intrinsic electronic phase separation arises. In this framework, 
we showed that the instability occurs for reasonable values of the Rashba coupling and the band 
structure parameters.  We then systematically analyzed how the phase separation instability changes 
upon varying the gating $V_g$, the temperature $T$ and the magnetic field $B$ (applied parallel to 
the interface). The dependence on the parameters defining the RSOC, $\beta$ and $\gamma$, and the 
band structure $\Delta$ and the heavy-to-light mass ratio $\nu$ is discussed in Appendix A, 
whereas the case of more realistic band structures is dealt with in Appendix B. 
Also the case of a $k$-cubic RSOC is addressed in Appendix C, where a coupling is considered leading to a
band splitting of the same order of the one obtained in Ref. \onlinecite{held}.

In all the above cases, a main finding  of our work is the possible occurrence of the phase separation instability 
for quite realistic values of the RSOC as soon as the heavier (mostly with $d_{xz,yz}$ 
character) bands start to be filled.
The filling of these bands, despite their location at higher energies $\Delta\sim 0.05$ eV, is a 
reasonable occurrence at the accessible electron densities both on experimental\cite{salluzzo} 
and theoretical grounds.\cite{parkmillis,sangiovanni,held,nayak}
Furthermore, the idea that qualitatively new physics then emerges when the
heavier $d_{xz}$ and $d_{yz}$ bands start to be filled is also gaining consensus on the basis of both 
theoretical results\cite{fischer} and experimental evidences.\cite{ilani,joshua,gabay,banerjee} 
Our scenario does not contradict these evidences, but suggests that these new physical effects may 
occur on an intrinsically inhomogeneous state.

In our opinion, a negative compressibility like the one measured in Ref. 
\onlinecite{mannhart} cannot be explained by impurities, defects and so on, but is a distinct 
signature of an intrinsic mechanism leading to an effective charge segregating attraction like the 
one proposed here.\cite{notamannhart}

In any case, the instability gives rise to dome-shaped coexistence surfaces in the parameter spaces
below which the electronic reconstruction due to the polarity catastrophe and the related electron 
densities cannot occur homogeneously, but rather give rise to domains with different densities. The 
Maxwell construction only determines the densities of the coexisting phases, but says nothing on the 
size and structure of the phase separated regions. These non universal details are instead determined 
by the energy cost of the interfaces between regions with different densities, by the strain effects, 
and so on. The residual amount (if any) of Coulombic forces opposing phase separation, also contributes 
to non universal features like size and shape of the inhomogeneous regions.\cite{emerykivelson,FPS}

Another remarkable finding is related to the closing of the phase-separation domes upon varying the 
gating potential. This gives rise to quantum critical points of a novel type, where the Landau damped 
electron density fluctuates with a relatively large dynamical critical index $z=3$,\cite{CDG,maccarone} 
or $z=4$. The fact that (nearly) neutral large-scale density fluctuations might occur in these systems 
is the key point allowing to escape the Coulombic frustration usually invoked to produce $z=1$ critical 
fluctuations.\cite{efetov,fisher,vishwanath} Of course, the possibility is also open that Coulombic 
forces frustrate the phase separation giving rise to charge-density waves (or even anharmonic stripes) 
similar to what has been proposed in high $T_c$ 
cuprates,\cite{emerykivelson,FPS,CDG,sulpizi,DBC,seibold2000,mazza} likely yielding a dynamical critical 
index $z=2$.\cite{CDG}

Recent experiments\cite{doublecrit} do reveal the presence of quantum critical behavior under magnetic 
field in the low voltage region, suggesting non trivial interplay between the superconducting fluctuations
and the nanoscale inhomogeneities. Work is in progress to investigate the possible connection between
these observed quantum critical behaviors and the quantum critical points found in the present work.

The generality of our model is based on the idea that RSOC arises at interfaces where a metallic 2DEG forms 
with a density which self-consistently adjusts itself on the basis of a local confining electric field. If 
one allows the RSOC to depend on this density via the orthogonal electric field, the issue of a charge 
instability naturally arises. Of course, the details of the system then enter to establish whether or not 
the frustration due to long-range Coulombic forces is present or not. It is anyhow tempting to investigate 
whether our instability mechanism is applicable to other systems as well. For instance, one could 
consider an EG with a sizable RSOC and/or very low densities in quantum wells, at the
boundaries of heavy metal alloys like Bi$_x$Pb$_{1-x}$, on the reconstructed surfaces of Ag(111), in MOSFET 
and semiconducting heterostructures at low densities, and on the surface of topological insulators.

Previous work in strongly correlated systems\cite{CDG,GC,GRCDK,BTDG,BKCDG,CCCDGR,CDG1} 
shows that electron-electron correlations and electron-phonon coupling favor phase separation. Therefore, 
while it is natural that weak repulsive electron-electron interactions turn out to weakly stabilize the 
phase-separation instability,\cite{finocchiaro}  strong correlations induce an intrinsic non-rigidity to 
the quasiparticle bands and may cooperate with the RSOC to produce a charge instability. 

On more general grounds, the present analysis paves the way for further models apt to describe the 
inhomogeneities at the LXO/STO interface. From experiments and theory we know that the superconducting 
state is inhomogeneous.\cite{BCCG,caprara} As a possible variation/improvement of our approach, one 
could then think of a model with superconductivity in which the coupling constant between Cooper pairs 
(or even superconducting islands) depends on the density. Based on the current work, one expects that 
also in this case the effective attraction leads to an unstable EG. Of course, one would need to 
check whether the orders of magnitude of the involved physical quantities are such to cause a 
measurable instability or not.

\vskip 0.5truecm 
\par\noindent 
{\bf Acknowledgments.} 
We acknowledge insightful discussions with L. Benfatto, N. Bergeal, C. Castellani, C. Di Castro, 
J. Lesueur, and R. Raimondi. G.S. acknowledges support from the Deutsche Forschungsgemeinschaft. M.G. 
and S.C. acknowledge financial support from University Research Project of the University of Rome 
Sapienza, No. C26A125JMB.

%%%%%%%%%%%%%%%% appendix %%%%%%%%%%%%%%%%%
\begin{appendix}
\section{How the instability depends on the parameters of the RSOC and of the bands}
\begin{figure}[h]
\begin{center}
\includegraphics[width=1.11\linewidth]{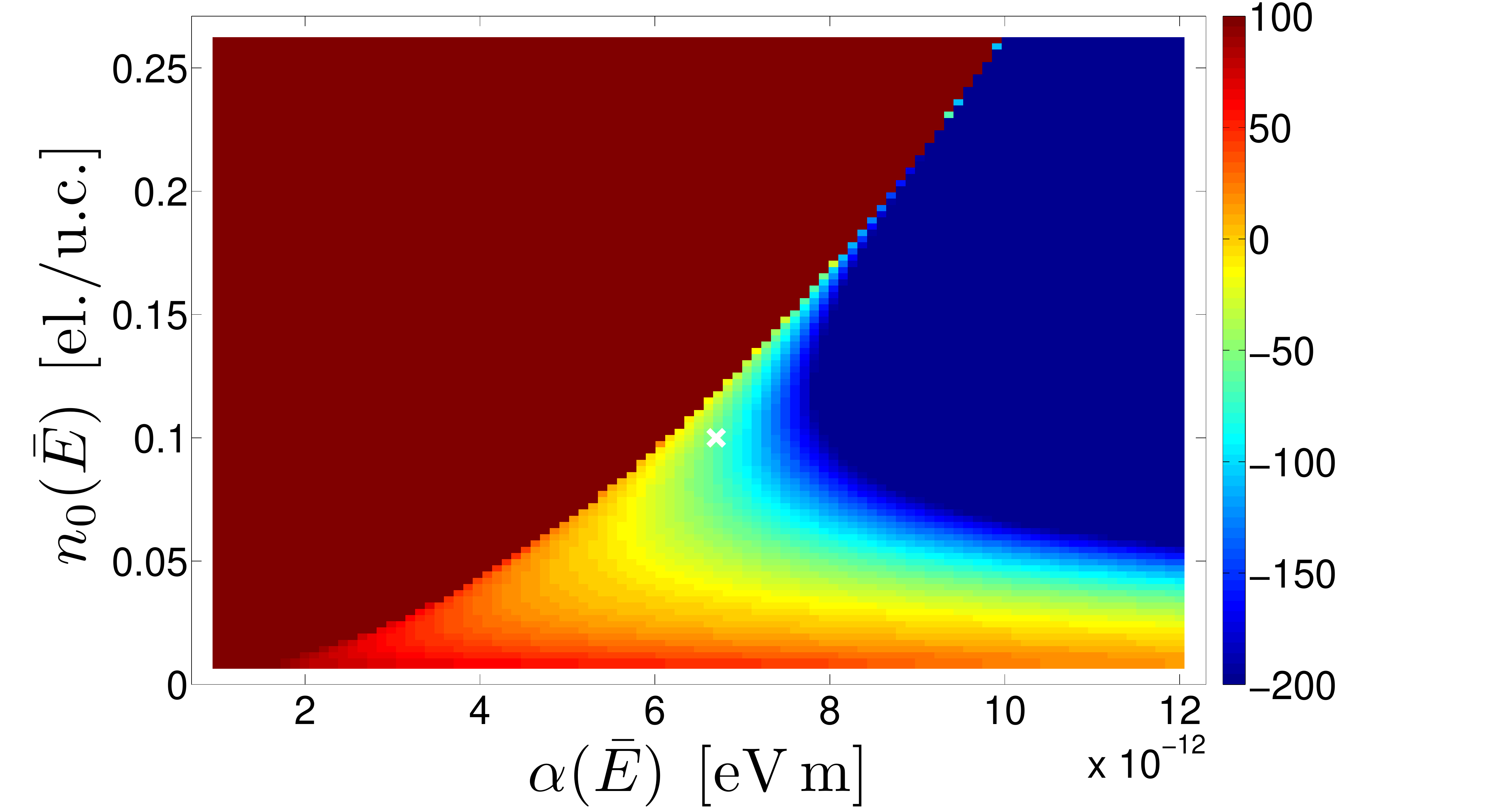}
\caption{(color online) Critical gate voltage $V_{gc}$ as a function of $\alpha(\bar{E})$ and 
$n_0(\bar{E})$.
The white cross corresponds to the values of $\alpha(\bar{E})$ and $n_0(\bar{E})$ chosen throughout 
the paper.}
\label{Loop_over_alpha_beta}
\end{center}
\end{figure}

We have shown the rather delicate dependence of the phase instability on the values of $\alpha$ 
and $\alpha'$. This suggests a strong dependence on the parameters $\beta$ and $\gamma$ as well. 
To investigate in more detail this dependence, we solve Eq. (\ref{dmu}) for a range of values of 
$\beta$ and $\gamma$ and determine the minimum gate voltage at which the instability occurs. The 
possible case of a phase separation only in the isotropic band is taken into account by calculating 
$\mu'$ for $n_0<\bar{n}_0$ from Eqs. (\ref{musi}) and (\ref{lambdai}) versus 
the density. In this way we get an understanding about how much the quantum critical point $
V_{gc}$ changes upon varying the parameters of the model.
In Fig.  \ref{Loop_over_alpha_beta} we plot $V_{gc}$ as a function of $\alpha(\bar{E})\propto\gamma$ and 
$n_0(\bar{E})\propto\beta^{-1}$ which denotes the density $n_0$ at which $\alpha$ attains its maximal value. 
The values chosen correspond to a cut-off field $\bar{E}\in[5\times10^7,2\times10^9]$\,V\,m$^{-1}$ and a 
maximal Rashba coupling $\alpha(\bar{E})\in[10^{-12},1.2\times10^{-11}]$\,eV\,m. The possible interval 
of the quantum critical point is chosen as $V_{gc}\in[-200,100]$\,V. The dark red region (upper 
left corner) corresponds to a (possible) quantum critical point above $100$\,V where our 
quadratic approximation of the anisotropic band structure does not hold anymore, while the dark blue 
region (upper right corner) corresponds to values of $\beta$ and $\gamma$ such that $V_{gc}\leq-200$\,V.

For small densities $n_0(\bar{E})\lesssim0.02$\,el./u.c. the function $\alpha(n_0)$ attains its maximum 
in the isotropic band (unless we partially fill the band by gating) and the occurrence of a phase 
separation either requires a large $\alpha(\bar{E})$ to drive the instability in the isotropic band 
already, or additional electrons $\delta n$ to reach
$n_0(\bar{E})$ when $\mu$ has already entered the anisotropic bands.
In the former case, which occurs for $\alpha(\bar{E})\gtrsim8\times10^{-12}$\,eV\,m and 
$n_0(\bar{E})\lesssim0.01$\,el./u.c., we obtain $V_{gc}\sim0$\,V, whereas
in the latter case the critical voltage is necessarily shifted towards positive gating.
Increasing $n_0(\bar{E})$ the voltage necessary 
to partially fill the band decreases, and thus $V_{gc}$ as well. For large values of $\alpha(\bar{E})$ the 
minimal value $-200$\,V is reached at $n_0(\bar{E})\gtrsim0.05$\,el./u.c. and stays low all the way up to 
$n_0(\bar{E})\lesssim 0.25$\,el./u.c., where the critical value passes in an abrupt manner from 
$V_{gc}<-200$\,V 
to $V_{gc}>100$\,V.

\begin{figure}[h]
\begin{center}
\includegraphics[width=1.05\linewidth]{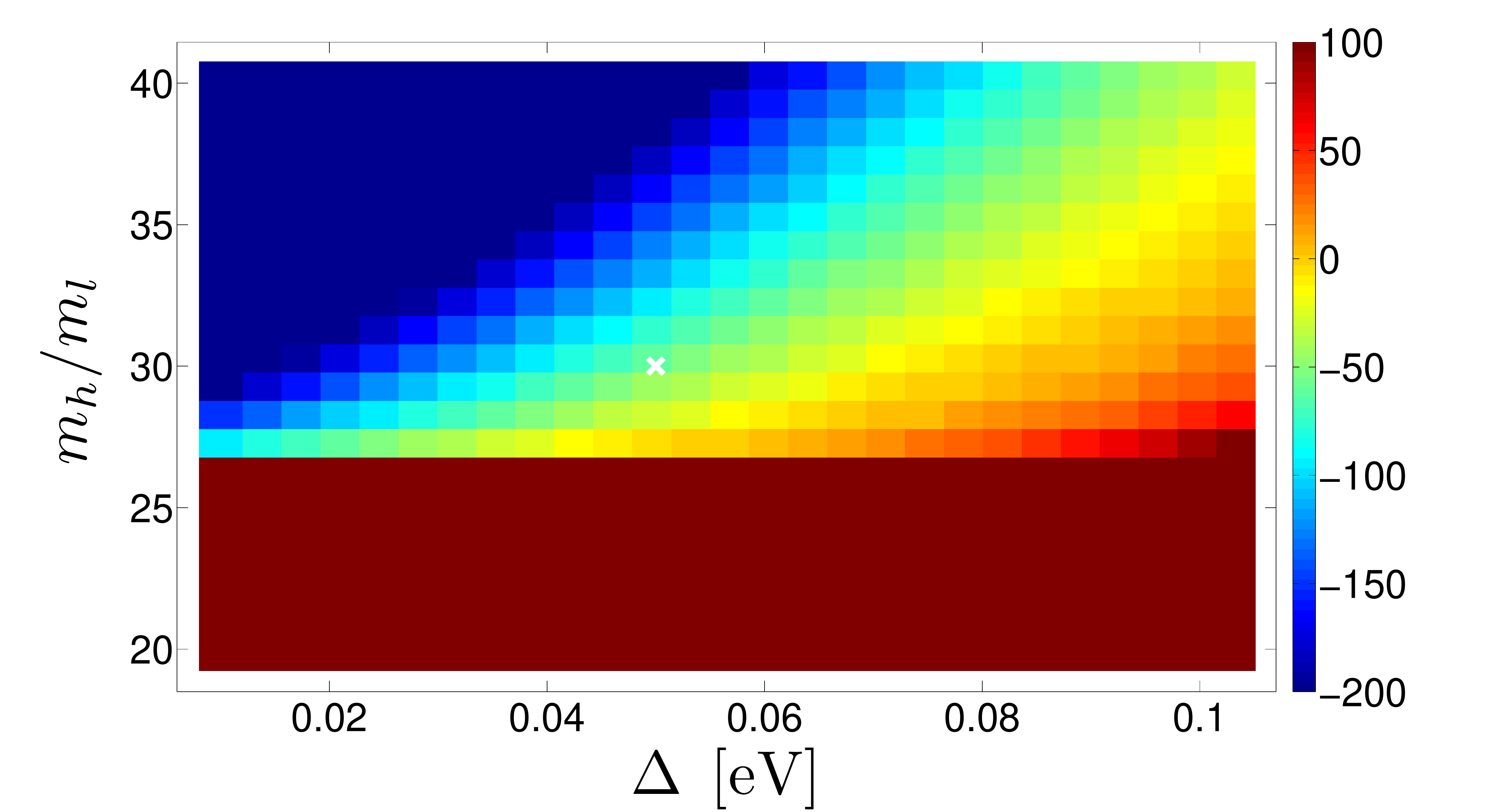}
\caption{(color online) Critical gate voltage $V_{gc}$ as a function of $\Delta$ and $\nu$. The white 
cross corresponds to the values of $\Delta$ and $\nu$ chosen throughout the paper.}
\label{Loop_over_Delta_nu_2}
\end{center}
\end{figure}

In addition to the parameters $\beta$ and $\gamma$ defining the RSOC, the model contains the band 
structure parameters $\Delta$ and $\nu=m_h/m_l$, where $m_l=0.7\,m_0$.
To gain insights about the dependence of $V_{gc}$ on these two parameters, we solve Eq. (\ref{dmu}) for 
$\Delta\in
[0.01,0.1]$\,eV and $\nu\in[20,40]$ with $\beta$ and $\gamma$ the same as initially 
($\beta=8.45\times10^{-10}$\,m\;V$^{-1}$ and 
$\gamma\equiv27\alpha(\bar{E})\beta/4=2.28\times10^{-20}$\,eV$\,$m$^2$$\,$V$^{-1}$). 

The resulting phase diagram reported in Fig. \ref{Loop_over_Delta_nu_2}
shows that for small values of the gap $\Delta\sim0.01$\,eV the instability arises at low 
gating because the chemical potential enters soon in the anisotropic bands. Then, 
increasing the gap leads to an increasing $V_{gc}$ because one has to introduce 
more electrons in the system before the anisotropic bands start to be filled.

Concerning the masses, the ratio has to exceed $\nu=26$ for the the EG to become unstable
in the physically meaningful range of gate voltages. 
This value corresponds to a large heavy mass of $18m_0$, especially when compared with 
the values reported from DFT calculations. Roughly speaking, if the mass is reduced by a factor of $x$, 
the RSOC has to be increased by a factor of $x$ in order to get the same effect. More to the point, 
consider the Hamiltonian
\begin{eqnarray}
{\cal H}&=&\frac{\hbar^2k^2}{2m}\pm\alpha k\label{H1}\\
&=&\frac{1}{x}\Biggl[\frac{\hbar^2k^2}{2m/x}\pm\alpha x k\Biggr]\label{H2}.
\end{eqnarray}
where Eq. (\ref{H2}) leads to the same energy spectrum as Eq. (\ref{H1}) (up to a factor $x$).

\begin{figure}[h]
\begin{center}
\includegraphics[width=1.\linewidth]{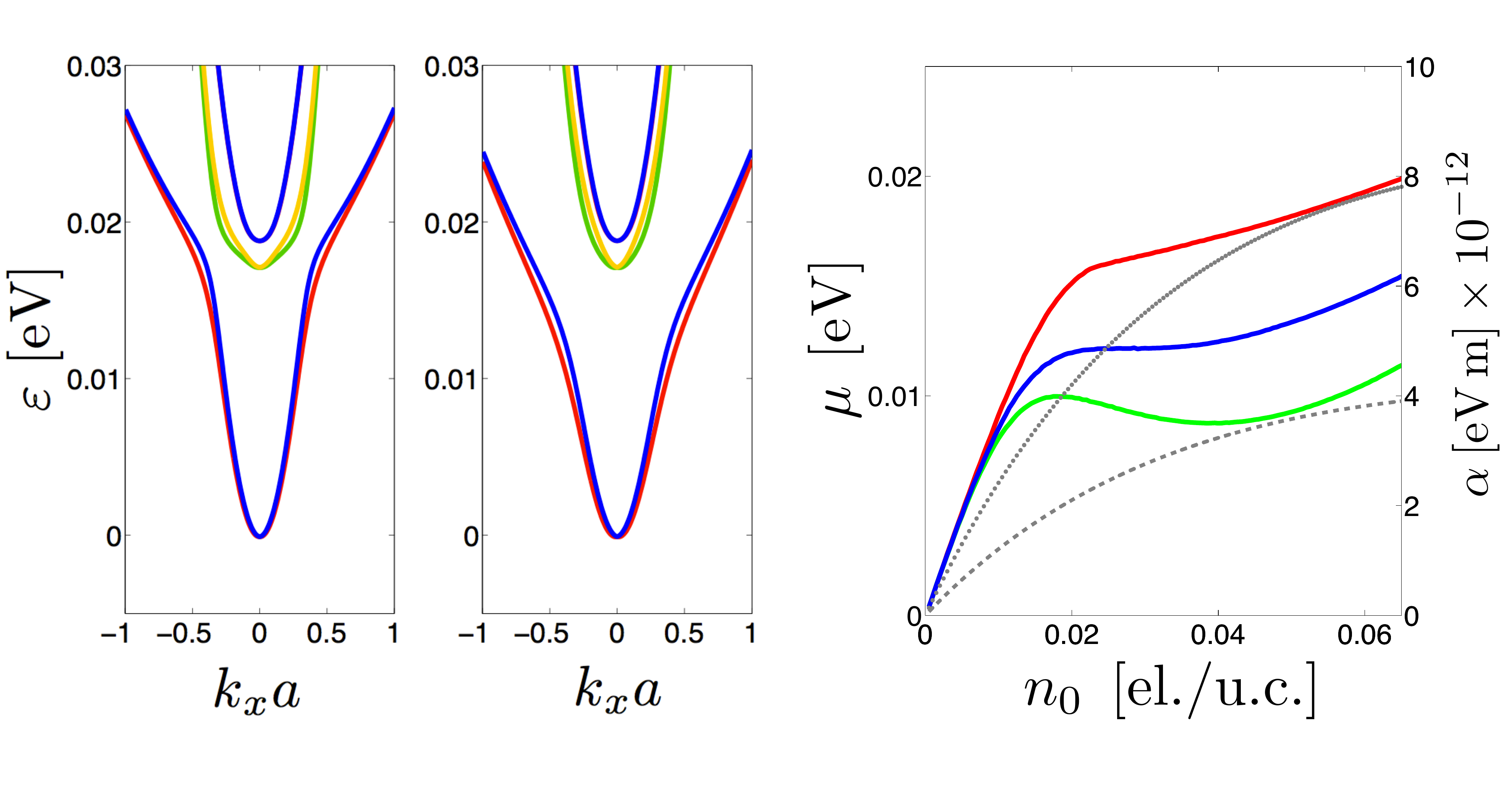}
\caption{(color online) Band structure for $n_0=0.025$\,[el./u.c.] with $\alpha(n)$ according to Eq.
(\ref{alpha_n0}) and parameters $\xi=0.005$ meV, $\beta=8.45\times10^{-10}$\,m\;V$^{-1}$, 
$\alpha^{max}=0.41\times10^{-11}$\,eV$\,$m (left panel) $0.82\times10^{-11}$\,eV$\,$m (middle panel).
Right panel: chemical potential $\mu(n_0)$ for parameters $\xi=0.005$ meV, 
$\beta=8.45\times10^{-10}$\,m\;V$^{-1}$, and $\alpha^{max}=0.41\times10^{-11}$\,eV$\,$m (red curve), 
$\alpha^{max}=0.67\times10^{-11}$\,eV$\,$m (blue curve),
and $\alpha^{max}=0.82\times10^{-11}$\,eV$\,$m (green curve).}
\label{Bandstructure_and_mu_vs_n_6by6}
\end{center}
\end{figure}

\section{More realistic band structures}
In the present appendix, we show that our conclusions concerning RSOC-driven electronic phase
separation also hold for more realistic band structures, which
we borrow from a tight-binding transcription of the results of DFT band calculations. Specifically, 
we numerically solved a tight-binding Hamiltonian $H=H_0 + H_{SO}$ in the form of a $6\times 6$ matrix 
in the space of the three $t_{2g}$ orbitals of Ti and spin.
$H_0$ describes the interface hopping and splitting in the $(d_{xy},d_{xz},d_{yz})$ basis, and is
diagonal in spin space (i.e., its spinor structure is proportional to the $2\times 2$ identity
matrix $\sigma_0$),
\begin{widetext}
\begin{equation*}\nonumber
H_0=\left (
\begin{array}{ccc}
-2t_1  \left( C_x  +  C_y\right)   -4t_3 \left( C_{xy}-1\right)&  0     & 0   \\
0     & -2t_1 C_x   -2t_2C_y  -2t_3C_x+\Delta  & 0 \\
0 & 0 &   -2t_1 C_y   -2t_2C_x  -2t_3C_y+\Delta 
\end{array}
\right )\otimes \sigma_0,
\end{equation*}
where $C_{x,y}\equiv \cos(k_{x,y})-1$, $C_{xy}=\cos(k_x)\cos(k_y)$, $t_1=0.277$ eV is the larger 
nearest neighbor  hopping corresponding to the direction of lighter band mass, $t_2=0.031$ eV is 
the  smaller nearest neighbor  hopping corresponding to the direction of heavier mass, while 
$t_3=0.05$ eV is the planar next-nearest neighbor hopping. The spin-orbit Hamiltonian
reads
\begin{equation*}\nonumber
H_{SO}=\left (
\begin{array}{cccccc}
0 &  0     &  -2i\alpha S_y & i\xi/2 & 0 & \xi/2  \\
0     & 0 &  i\xi/2 &  -2i\alpha S_y & \xi/2 & -2i\alpha S_x  \\
2i\alpha S_y  & -i\xi/2 &  0 & 0 &  i\xi/2 &0 \\
0 & 2i\alpha S_y  & 0 & 0 & 0 & - \xi/2 \\
2i\alpha S_x & \xi/2 & -\xi/2 & 0 & 0 \\
-\xi/2 & 2i\alpha S_x & 0 & i \xi/2 & 0 & 0
\end{array}
\right ),
\end{equation*}
\end{widetext} 
where $S_{x,y}\equiv \sin (k_{x,y})$, $\xi=2\Delta^{SO}/3$ is the atomic spin-orbit strength (see Ref.
\onlinecite{held}), while $\alpha=\alpha(n_0)$ is the inter-orbital hopping term due to the interface 
asymmetry, eventually producing the Rashba-like band splittings (see Refs. 
\onlinecite{held}, \onlinecite{nayak}, and \onlinecite{khalsa} for details). The resulting band 
structures are reported in the left 
and middle panel of Fig. \ref{Bandstructure_and_mu_vs_n_6by6}. For the parameters of $\alpha(n_0)$ 
we phenomenologically adjusted $\beta $ and $\gamma$, such as to yield a value 
$\alpha(0.025)=2-4\times10^{-12} $ eV\,m  at $n_0=0.025$ el./u.c. The value $\xi=0.005$ eV is 
accordingly adjusted to produce a band 
structure similar to the one reported in Fig. 3 of Ref. \onlinecite{nayak}. The chemical potential 
as a function of $n_0$ is the red curve reported in the right panel of Fig. 
\ref{Bandstructure_and_mu_vs_n_6by6}. For this choice of the parameters, the system is still stable, 
although a substantial decrease of the inverse compressibility (the slope of the
$\mu$ vs. $n_0$ curve) is found.
Remarkably, by simply increasing by a factor two the strength of 
$\alpha$, we find a negative compressibility. We conclude that also for more complex (and 
realistic) band structures the instability occurs in a range of RSOC strengths which is fully 
compatible with the experimentally inferred values. 
This result is quite reasonable. Indeed, although the band structure of Fig. 
\ref{Bandstructure_and_mu_vs_n_6by6} looks quit different form the bands described by 
Eqs. (\ref{disp_rel_iso}) and (\ref{disp_rel_aniso}), it results in fact from an admixture and 
spin-orbit splitting of the $(d_{xy},d_{xz},d_{yz})$ bands, which were split by the RSOC, 
but not mixed (by hybridization and/or spin-orbit coupling), within our simplified approach. 
For similar values of the physical parameters, the two band structures qualitatively share the 
same physics, with minor quantitative differences. On the other hand, the simplified band structure 
adopted in the main part of this piece of work has the advantage of allowing for more transparent 
analytical results, highlighting the physical mechanisms leading to RSOC-driven electronic 
phase separation.

\section{Cubic Rashba coupling}
\begin{figure}[h]
\begin{center}
\includegraphics[width=1.\linewidth]{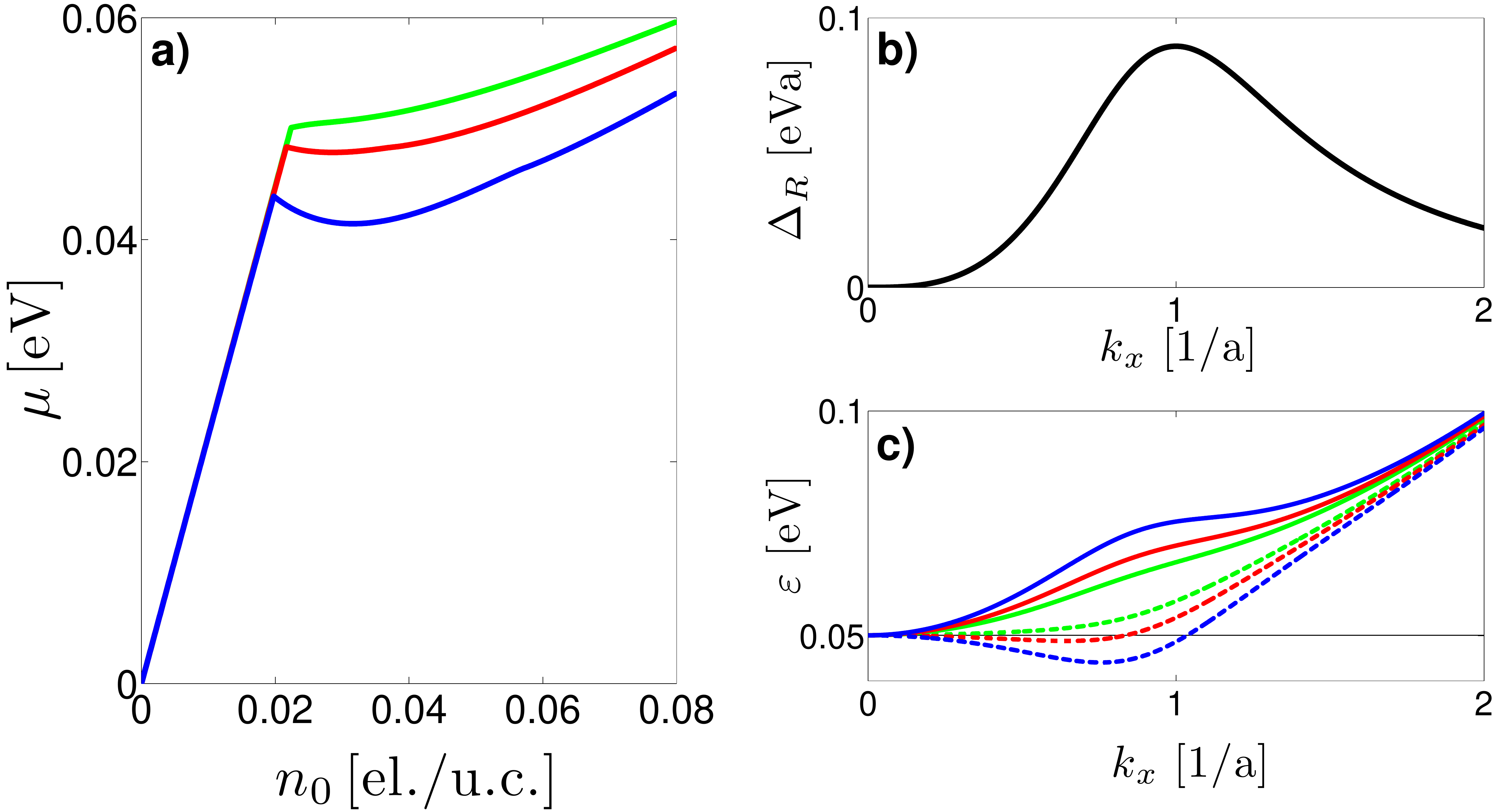}
\caption{(color online) (a) Chemical potential as a function of $n_0$ for three values of $\alpha_3$.
$\alpha_3=0.67\times 10^{-11}$ eV\,m (green line), $\alpha_3=1.33\times 10^{-11}$ eV\,m (red line), 
$\alpha_3=2.0\times 10^{-11}$ eV\,m (blue line). 
(b) Plot of the momentum dependent RSOC factor $\Delta_{R}(k)$. (c)
Band structure for $n_0=0.02$\,[el./u.c.] with $\alpha(n_0)$ according to Eq.
(\ref{alpha_n0})}
\label{Rashba_RESUB_Fig17}
\end{center}
\end{figure}

For the sake of completeness, in this Appendix we briefly consider the possibility that RSOC lacks $k$-linear dependence
and starts with a $k^3$ term\cite{nayak} in the anisotropic bands.  
To implement this RSOC we phenomenologically start from the simplified expression of the band 
structure in Eqs.(\ref{acca}) and (\ref{H_Rashba}) and take a momentum dependent spin splitting of the form
\begin{equation}\label{kcubic}
\Delta_{R}(k)=2\,\frac{a^2k^3}{(ak)^6+1}.
\end{equation}
to replace the $k$-linear term ($a$ is the lattice spacing). The denominator has been chosen to match the generic shape of the
kubic spin splitting (see, e.g., Ref. \onlinecite{khalsa,held}).
The strength of this RSOC is varied by multiplying  $\Delta_{R}(k)$ by a factor
$\alpha_3(n_0)$, with the same density dependence as in the main part of the paper [see Eq. (\ref{alpha_n0}) and Fig.\ref{mu_n_T0}]
The resulting anisotropic band structure in the heavy-mass direction acquires a
dispersion as reported in Fig.\ref{Rashba_RESUB_Fig17}.

As it can be seen in Fig. \ref{Rashba_RESUB_Fig17} (a), a phase separation instability takes place also in this
case for values of $\alpha_3\approx 1.33\times 10^{-11}$ eV\,m, such that the lowering of the lower part of the
 (heavy) bands is comparable to the analogous lowering in the case of $k$-linear RSOC. Indeed the chemical potential
 starts to bend down with a negative slope as soon as the Fermi energy enters the bottom of the band provided 
 this is located at finite momenta, where the momentum dependent RSOC has a finite value.
 
 Notice that the chosen values of the parameters for the RSOC lead to a band splitting of the order of 25 meV, comparable with the
 values obtained in Ref. \onlinecite{held}.
\end{appendix}


\begin{thebibliography}{99}
\bibitem{notapotential} This assumption holds exactly for a triangular confining 
potential, while it becomes questionable for electron sub-bands, which are less confined in the
$z$ direction because the potential bends and deviates from linearity far from
the interface.
\bibitem{notamannhart} Several important differences make a precise comparison with the experimental 
data of Ref. \onlinecite{mannhart} difficult. First of all the top-gating used in experiments act on 
the filling differently from the back-gating more commonly considered both in experimental set-ups 
and in models for the polarity catastrophe. Secondly the experimental inverse compressibility values 
greatly depend on the samples, making a quantitative comparison hard to implement. Finally, some 
physical effects like electron-electron interaction and disorder become more important at low density 
($n\sim 10^{-3}$ [el/u.c.]), thereby preventing a reliable quantitative use of our schematic model.
\end{thebibliography}
\end{document}